\documentclass[showpacs,showkeys,amsmath,amssymb,aps,rmp]{revtex4-1}
\usepackage{graphicx}
\usepackage{dcolumn}
\usepackage{bm}

\begin{document}

\title{Molecular machines operating on nanoscale: from classical to quantum}
\author{Igor Goychuk}
\email{igoychuk@gmail.com}
\affiliation{Institute for Physics and Astronomy, University of Potsdam, 
Karl-Liebknecht-Str. 24/25, 14476 Potsdam-Golm, Germany}

\begin{abstract}
The main physical features and operating principles of isothermal
nanomachines in microworld are reviewed, which are common for both classical and quantum
machines. Especial attention is paid to the dual and constructive 
role of dissipation and thermal fluctuations,
fluctuation-dissipation theorem, heat losses and free energy transduction, thermodynamic
efficiency, and thermodynamic efficiency at maximum power. Several basic models are
considered and discussed to highlight generic physical features. Our exposition allows
to spot some common fallacies which continue to plague  the literature, in particular,
erroneous beliefs that one should minimize friction and lower the temperature to arrive
at a high performance of Brownian machines, and that thermodynamic efficiency at maximum
power cannot exceed one-half. The emerging topic of anomalous molecular motors operating 
sub-diffusively but highly efficiently 
in viscoelastic environment of living cells is also discussed. 
\end{abstract}

\maketitle

\section{Introduction}

Myriad of miniscule unseen in standard classical optical microscopes molecular 
nanomotors operate in living cells doing various tasks while utilizing  metabolic energy,
stored e.g. in 
ATP molecules maintained at out-of-equilibrium concentrations, or in nonequilibrium ion 
concentrations across biological membranes.  And also vice versa they replenish 
the reserves of metabolic energy using other
sources of energy, e.g. light by plants, or energy of covalent bonds of various food molecules
by animals \cite{Pollard08}. The main physical principles of their operation are more or less
understood by now \cite{Hill,Nelson}, even if hardly one can state at present that the work
of a single particular molecular motor, e.g. a representative of a large family
of kinesin motors is understood and described in all finest statistico-mechanical detail.
The advances and perspectives of nanotechnology compel us humans to devise our own
nanomotors \cite{Kay,Erbas,Cheng}. We do believe that learning from nature can 
help to make the artificial
nanomotors highly efficient, and maybe even better than natural. 
On this way, to understand the main physical operating
principles within the simplest minimalist physical models can help a lot indeed.

First of all, any periodically operating motor or engine requires a working body
undergoing cyclic changes and a source of energy to drive such cyclic changes.
Furthermore, it should be capable to do a useful work on external bodies.
In case of thermal heat engines, the source of energy is provided by a heat exchange with two
heat reservoirs or bathes at different temperatures $T_1$, and $T_2>T_1$ with the maximally possible
Carnot efficiency $\eta_C=1-T_1/T_2$ \cite{Callen}. This very famous textbook result of classical 
thermodynamics or, better, thermostatics is to be modified, when the heat flows are 
considered in time. Then, for infinitesimally slow heat flows occurring yet in finite time, one 
obtains the  Curzon and Ahlborn result $\eta_{CA}=1-\sqrt{T_1/T_2}\leq \eta_C$ \cite{Callen,Curzon}. 
The analogy
with heat engines is, however, rather misleading for isothermal engines operating
at the same temperature, $T_1=T_2$. Here, the analogy with electrical motors is much better.
It becomes almost literal in the case of rotary ATP-synthase \cite{Yoshida} or flagellar bacterial motors 
-- the electrical nanomotors
of living cells, -- where the energy of 
proton electrochemical gradient (an electrochemical rechargeable battery) is used to 
synthesize ATP molecules out of ADP and the orthophosphate $\rm P_i$ (the useful work done),
in the case of ATP-synthase, or to produce mechanical motion by flagellar motors 
\cite{Pollard08,Nelson}. 
Such a nanomotor as ATP-synthase can operate also in reverse \cite{Yoshida}, 
and the energy of ATP hydrolysis can be used to pump protons against their electrochemical
gradient to recharge ``the battery''. Such and similar nanomotors can operate with 
isothermal thermodynamic efficiency, defined as the ratio of useful work done to the
input energy spent, close to one, at ambient temperatures in a highly
dissipative environment. This is a first counter-intuitive remarkable feature, which needs to
be explained. It is easy to derive this result within a simplest model (see below)
for an infinitesimally slow operating motor, at zero power. At the maximum power, at a finite
speed, the maximal thermodynamic efficiency within such a model is one-half. This
is still believed by many to be the maximally possible thermodynamic efficiency of isothermal
motors at maximum power, in principle, as a theoretical bound. 
However, this belief born in underestimating the role played by thermal
fluctuations in nonlinear stochastic dynamics 
and the role of fluctuation-dissipation theorem or FDT on nano- and microscale, 
is generally wrong. It is valid only for some  particular dynamics, as we shall clarify below
giving three counter-examples to it.
The presence of strong thermal fluctuations at ambient temperatures, playing a constructive
and useful role, is a profound physical feature of nanomotors, as compare with macroscopic
motors of our everyday experience. One requires to understand and to develop intuition for this
fundamental feature. Nanomotors are necessarily Brownian engines, very differently
 from macroscopic ones.

\section{Fluctuation-dissipation theorem, the role of thermal fluctuations}

Motion in any dissipative environment is necessarily related to dissipation of energy. Particles
experience a frictional force, which in a simplest case of Stokes friction is linearly proportional
to the particle velocity with a viscous friction coefficient which will be denoted as $\eta$. If the
corresponding frictional energy losses would not be compensated by an energy supply, the motion
would eventually stop. However, this never happens in microworld for micro- or nanosized particles. 
Their stochastic Brownian motion can persist forever even at thermal equilibrium. The energy 
necessary for this is supplied by thermal fluctuations. Therefore, friction and thermal noise are
intimately related which is a physical context of fluctuation-dissipation theorem \cite{Kubo66}.
Statistical mechanics allows to develop a coherent picture to rationalize this fundamental feature
of Brownian motion.
We start with some generalities which can be easily understood within a by now standard dynamical
approach to Brownian motion that can be traced back to pioneering contributions by 
Bogolyubov \cite{Bogolyubov},
Ford, Kac, Mazur \cite{Ford65,Ford88}, and others. 
Consider a Brownian particle with mass $M$, coordinate $x$, and
momentum $p$. It is subjected to a regular dynamical force $f(x,t)$, as well as
 frictional and stochastically fluctuating forces
of the environment. These later ones are modeled by an elastic coupling of 
this particle to a set of $N$ harmonic oscillators with masses $m_i$, coordinates $q_i$,
and momenta $p_i$. We take this coupling in the form 
$V_{\rm int}=\sum_{i=1}^N \kappa_i(x-q_i)^2/2$, with spring constants
$\kappa_i$.
This is a standard mechanistic model of nonlinear classical Brownian motion 
 known within quantum dynamics also as Caldeira-Leggett model \cite{Caldeira83} upon
a modification of the coupling term, or making a canonical transformation \cite{Ford88}. 
 Both classically and quantum-mechanically \cite{Ford88} (in Heisenberg picture) the equations
of motion read
\begin{eqnarray}
\dot x &= &p/M, \nonumber \\
\dot p & = & f(x,t)- \sum_{i=1}^{N} \kappa_i (x-q_i), \label{A1}\\
\dot q_i &= &p_i/m_i, \nonumber\\
\dot p_i & = & \kappa_i (x-q_i)\;. \label{A2}
\end{eqnarray}
In quantum case, $x$, $q_i$, $p$, $p_i$ are operators obeying commutation relations
$[x,p]=i\hbar$, $[q_k,p_j]=i\delta_{kj}\hbar$, $[x,q_i]=0$, $[p,p_i]=0$. Force $f(x,t)$
is also operator. 
Using Green function of harmonic oscillators
the dynamics of bath oscillators can be excluded (projection
of hyper-dimensional dynamics on (x,p) plane) and presented further merely by the initial 
values $q_i(0)$ and $p_i(0)$. This results in a Generalized Langevin
Equation (GLE) for the motor variables
\begin{equation}\label{GLE}
M\ddot x+\int_{0}^{t}\eta(t-t')\dot x(t')dt'=f(x,t)+\xi(t),
\end{equation}
where
\begin{eqnarray}\label{kernel}
\eta(t)=\sum_{i}\kappa_i\cos(\omega_i t),
\end{eqnarray}
is a  memory kernel and
\begin{eqnarray}\label{noise}
\xi(t)=\sum_i \kappa_i \left ( [q_i(0)-x(0)]\cos(\omega_i t)+
\frac{p_i(0)}{m_i\omega_i}\sin(\omega_i t) \right ) \; 
\end{eqnarray} 
is a bath force, where $\omega_i=\sqrt{\kappa_i/m_i}$ are the frequencies of bath
oscillators.
Eq. (\ref{GLE}) is still a purely dynamical equation of motion which is exact.
 The dynamics of $[x(t),p(t)]$
is completely time-reversible for any given $q_i(0)$ and $p_i(0)$ by derivation,
unless the time-reversibility is  dynamically broken by $f(x,t)$, or by 
boundary conditions.
Hence, time-irreversibility within a dissipative Langevin dynamics is in the first line
a statistical effect due to averaging over many trajectories. Such an averaging cannot
be undone, i.e. there is no way to restore a single trajectory from
their ensemble average. Consider further first a classical dynamics.
Let us choose initial $q_i(0)$ and $p_i(0)$ from a canonical 
hyper-dimensional Gaussian distribution $\rho(q_i(0),p_i(0))$, zero-centered
in $p_i(0)$ subspace and centered around $x(0)$ in $q_i(0)$ subspace, and
characterized by the thermal bath temperature $T$,
like in a typical molecular dynamics setup.  Then, each
$\xi(t)$ presents a realization of a stationary zero-mean
Gaussian stochastic process which can be completely
characterizes by its autocorrelation function $\langle \xi(t)\xi(0)\rangle$.
Here, $\langle ... \rangle$ denotes statistical averaging done with $\rho(q_i(0),p_i(0))$.
An elementary calculation yields the fluctuation-dissipation relation (FDR)
named also the second FDT by Kubo \cite{Kubo66}:
\begin{equation}\label{FDR}
\langle \xi(t')\xi(t)\rangle =k_B T\eta(|t-t'|).
\end{equation} 
Notice that it valid even for a thermal bath consisting of a single oscillator.
However, a quasi-continuum of oscillators is required to render the random force
correlations decaying to zero in time. This is necessary for $\xi(t)$ to 
be ergodic in correlations.
Kubo obtained this FDT in a very different way, namely by considering the processes
of dissipation caused by phenomenological memory friction characterized by the memory
kernel $\eta(t)$, i.e. heat given by the particle to the thermal bath, 
and absorption of energy from 
the random force $\xi(t)$, i.e. heat absorbed from the thermal bath, so
that the both processes are balanced at thermal equilibrium, an the averaged 
kinetic energy of Brownian particle is $k_BT/2$, in accordance with equipartition theorem in classical
equilibrium statistical mechanics. This is a very important point.
At thermal equilibrium, the net heat exchange between the motor and its environment
is zero, for arbitrary strong dissipation. This is a primary and fundamental reason
why thermodynamic efficiency of isothermal nanomotors can in principle achieve unity,
in spite of a strong dissipation. 
For example, thermodynamic efficiency of F1-ATPase rotary motor can be close
to 100\% as recent experimental work implies \cite{Toyabe}.
For this to happen, the motor must operate most closely to thermal equilibrium, in order
to avoid net heat losses.
One profound lesson from this is that there is no need to minimize friction
on nanoscale, which (minimization of frictional losses) is  a very misleading 
misconception which continues to plague the research on Brownian motors -- the
 so-called dissipationless ratchets are worthless, see on this below. Very efficient
 motors can work at ambient temperatures, and arbitrary strong friction. 
 No need to go into a deep quantum cold
 which requires per se huge energy expenditure to create it in a lab.
 
Every thermal bath and its coupling to the particle 
can be characterized by the bath spectral density 
$J(\omega)=\frac{\pi}{2}\sum_i\frac{\kappa_i^2}{m_i\omega_i}\delta(\omega-\omega_i)=
\frac{\pi}{2}\sum_i
m_i\omega_i^3\delta(\omega-\omega_i)$ \cite{Caldeira83,Ford88,WeissBook}. It allows to express 
$\eta(t)$ as $\eta(t)=(2/\pi)\int_0^{\infty}d\omega J(\omega)\cos(\omega t)/\omega$
and the noise spectral density via  the Wiener-Khinchin theorem, $S(\omega)=
\int_{-\infty}^{\infty}\langle \xi(t)\xi(0)\rangle e^{i\omega t}dt$, as 
$S(\omega)=2k_BT J(\omega)/\omega$. The strict Ohmic model, $J(\omega)=\eta\omega$,
without frequency cutoff, corresponds to the standard Langevin equation
\begin{equation}\label{LE}
M\ddot x+\eta\dot x =f(x,t)+\xi(t),
\end{equation}
with uncorrelated white Gaussian thermal noise, 
$\langle \xi(t')\xi(t)\rangle =2 k_B T\eta \delta(|t-t'|)$. Such a noise is singular.
Its mean-square amplitude is infinite.
This is, of course, a very
useful but yet idealization.
A frequency cutoff must be physically present, which results in a thermal GLE description
with correlated Gaussian noise.

The above derivation can also be repeated straightforwardly  for quantum dynamics.
This leads to quantum GLE, which is looking formally the same as (\ref{GLE}) in Heisenberg picture
with the 
only difference: The corresponding random force becomes operator-valued with a complex-valued
autocorrelation function \cite{Ford88,Gardiner,WeissBook}
\begin{eqnarray}\label{LangevinQuantum_FDR}
\langle \hat \xi(t)\hat \xi(t')\rangle_{qm} & = &\frac{\hbar}{\pi}
\int_0^\infty d\omega J(\omega)\{\coth(\hbar \omega/2k_BT)  \\
&\times &\cos[\omega (t-t')] 
 -i\sin[\omega (t-t')]\}\;.  \nonumber
\end{eqnarray}
Here, the averaging is done with equilibrium density operator of the bath oscillators.
The classical Kubo result (\ref{FDR}) is restored in the formal limit $\hbar\to 0$.
To obtain a quantum generalization of (\ref{LE}), one can introduce a frequency cutoff,
 $J(\omega)=\eta\omega
\exp(-\omega/\omega_c)$, and split $\hat \xi(t)$ into a sum of
zero-point quantum noise $\hat\xi_{T=0}(t)$ and thermal quantum noise $\hat\xi_{T>0}(t)$
contributions, $\hat \xi(t)=\hat \xi_{T=0}(t)+\hat\xi_{T>0}(t)$.
This yields 
\begin{eqnarray}\label{qcorr}
\langle  \hat\xi(t) \hat\xi(0)\rangle_{qm} =\frac{1}{\pi}
\frac{\hbar\eta\omega_c^2(1-i\omega_c t)^2}{[1+(\omega_c t)^2]^2}+2k_BT \eta \delta_T(t)
\end{eqnarray}
with 
\begin{eqnarray}\label{quantum-T}
\delta_T(t)=\frac{1}{2\tau_T}\Big[\frac{\tau_T^2}{t^2}-\frac{1}{\sinh^2(t/\tau_T)} 
\Big],
\end{eqnarray}
where $\tau_T =\hbar/(\pi k_BT)$ is a characteristic time of thermal quantum fluctuations.
 Notice the dramatic change of quantum thermal 
correlations, from delta function at $\hbar\to 0$, 
to an algebraic decay $\delta_T(t)\sim t^{-2}$,
for finite $\tau_T$ and $t\gg \tau_T$. The total integral of $\delta_T(t)$ is unity, and
one of the real part of the $T=0$ contribution is zero.
In the classical limit, $\hbar\to 0$, $\delta_T(t)$ becomes delta-function.
Notice also that the first complex-valued term in Eq. (\ref{qcorr}), which corresponds to zero-point quantum 
fluctuations, starts from a positive singularity at the origin $t=0$
in the classical white noise limit, $\omega_c\to\infty$, and becomes  negative 
$-\hbar\eta/(\pi t^2)$ for $t>0$. Hence, it lacks a characteristic time scale. However,
it cancels precisely the same contribution, but with the opposite sign stemming
formally from the thermal part in the limit $t\gg \tau_T$, at $T\neq 0$. Thus, quantum
correlations, which correspond to the Stokes or Ohmic friction,
 are decaying in fact nearly exponentially for $\omega_c\gg 1/\tau_T$,
except for physically unachievable $T=0$.
Here, we see two profound quantum-mechanical features in the quantum operator-valued
 version
of classical Langevin equation (\ref{LE}) with memoryless Stokes friction:
First, thermal quantum noise is  correlated. Second, zero-point quantum
noise is present. This is the reason why quantum Brownian motion would not stop
even at absolute zero of temperature $T=0$. A proper treatment of these
quantum-mechanical features produced a large controversial literature in the case
of nonlinear quantum dynamics, whenever $f(x)$ is different from constant, or has a 
nonlinear dependence
on $x$,
see books \cite{Gardiner,WeissBook} for further references and details.
Indeed, dissipative quantum dynamics cannot be fundamentally Markovian, as
already our short exposition reveals. This is contrary to a popular approach 
based on the idea of quantum semi-groups which guarantees a complete  positivity
of such a dynamics \cite{Lindblad}. 
The main postulate of the corresponding theory (the semi-group property of the 
evolution operator expressing the Markovian character of evolution) simply
cannot be justified on a fundamental level, thinking in terms of interacting particles 
and fields
(a quantum field theory approach). Nevertheless, Lindblad theory and its allies,
for example stochastic Schroedinger equation \cite{WeissBook}, are extremely useful in quantum
optics where dissipation strength is very small. The applications to condensed matter
with appreciably strong dissipation should, however, be done with a great care.
They can lead to clearly incorrect results which contradict to exactly solvable models
\cite{WeissBook}.
Nonlinear quantum Langevin dynamics is very tricky, even within
a semi-classical treatment, where the dynamics is treated as classical but
with colored classical noise corresponding to the real part of $\hat \xi(t)$ treated
as c-number. 
As a matter of fact, quantum dissipative dynamics is fundamentally non-Markovian, which is a primary
source of all difficulties, and confusion.
Exact analytical results are  
practically absent (except for linear dynamics), and various Markovian approximations to nonlinear non-Markovian dynamics 
are controversial being restricted to
some parameter domains, e.g. weak system-bath coupling or a weak tunnel coupling/strong system-bath
coupling. Moreover, they are capable to easily produced unphysical results 
(like violation of the second law of thermodynamics) beyond their validity domains.

Furthermore,  a profoundly quantum dynamics has often just a few relevant
discrete quantum energy levels, rather than  a continuum of quantum states. 
Two-state quantum system serves as a prominent example. Here, one
can prefer a different approach to dissipative quantum dynamics, e.g. the reduced density operator 
method leading to quantum kinetic equations for level populations and coherences of 
the system of interest \cite{Nakajima58,Zwanzig60,Argyres64,GoychukAP05}.
It provides a description on the ensemble level and relates to quantum Langevin equation
in a similar manner as classical Fokker-Planck equation (ensemble description) relates
to classical Langevin equation (description on the level of single trajectories).

\subsection{Minimalist model of Brownian motor}

\begin{figure}
\includegraphics[width=16cm,keepaspectratio]{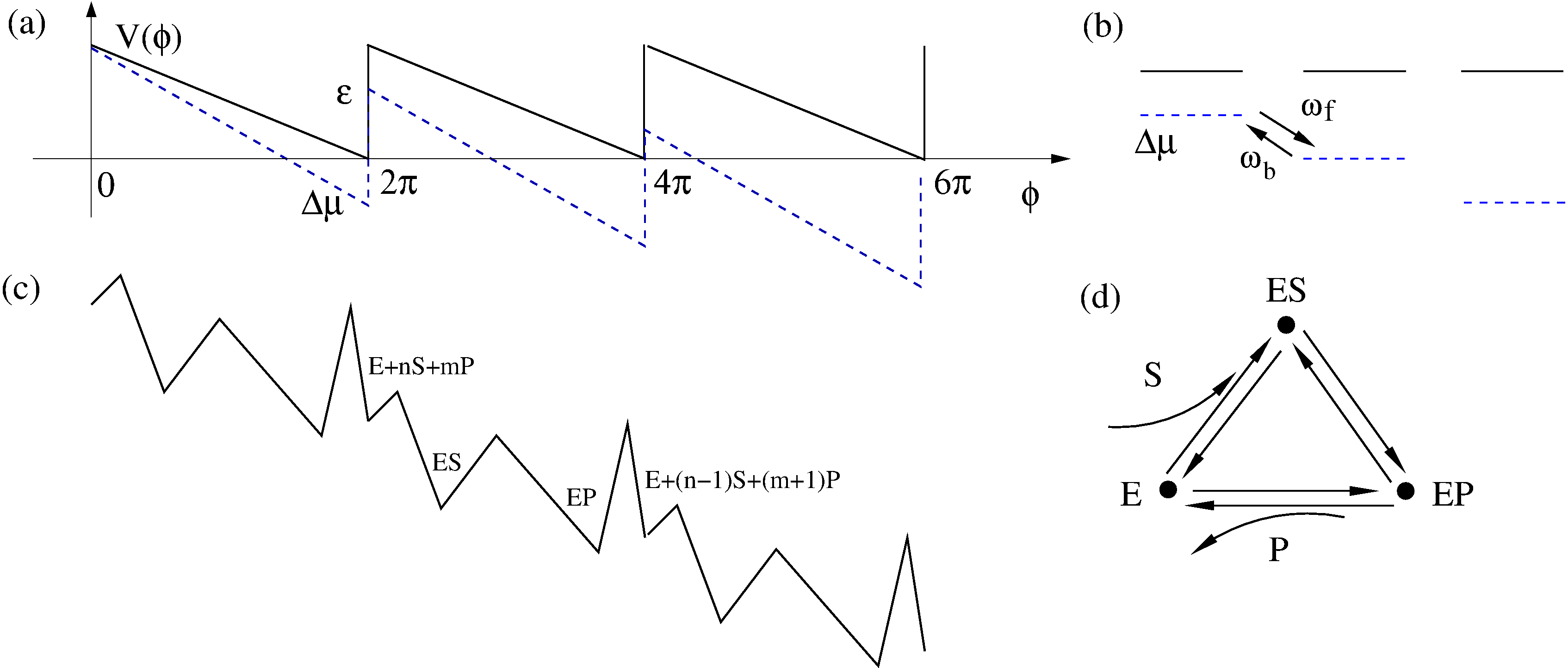}
\caption{(a) Simplest toy model for a periodic ratchet potential 
$V(\phi)$ with depth $\epsilon$.  
Bias $\Delta \mu<0$ per one rotation turn introduces directional rotations. 
(b) Discrete state model which corresponds to (a) with forward, $\omega_f$,
and backward,  $\omega_b$, rates calculated  e.g. by  solving Smoluchowski
equation, see the text.
This picture holds also quantum-mechanically with quantum mechanical effects
entering the rates in some models, where diagonal and off-diagonal elements of
reduced density matrix are completely decoupled in the energy basis of localized states
depicted.  
(c) The general
modeling route is inspired by enzyme dynamics, where an enzyme molecule cycles periodically between
substrate-free state \textbf{E}, state with bound substrate \textbf{ES} and state with 
bound product \textbf{EP}, which correspond to three metastable states of enzyme within a
continuum of conformational states. 
$\Delta \mu$ corresponds to the free energy released by transformation
$S\to P$ which drives the cyclic rotations of ``catalytic wheel'' 
\cite{Wyman,Rozenbaum,Qian}, see in (d).
 This energy can
be used to do a work against a loading force $f_L$, which is not shown.
For example, enzyme is an ion pump utilizing energy of ATP hydrolysis. ATP is substrate,
$\rm ADP+P_i$ is product. The useful work done is transfer an ion across a membrane 
against the corresponding electrochemical transmembrane gradient.}
\label{Fig1}

\end{figure}

A minimalist model of motor can be given by 1d cycling of motor particle 
in a periodic potential $V(\phi+2\pi)=V(\phi)$, see in Fig. \ref{Fig1},
 which models periodic turnovers of motor
within a continuum of intrinsic conformational states \cite{Nelson}. 
$\phi$ is a chemical cyclic reaction coordinate. 
The motor cycles can be driven
by an energy supplied by constant driving force, or rather torque $F$, with free 
energy $\Delta \mu=2\pi F$ spent
per one motor turn, and do a useful work against an opposing
torque or load $f_L$, so that the total potential energy is $U(\phi)=V(\phi)-F\phi+f_L \phi$.
Overdamped Langevin dynamics is described by
\begin{equation}\label{LE2}
\eta\dot \phi =f(\phi)+F-f_L + \xi(t),
\end{equation}
where $f(\phi)=-\partial V(\phi)/\partial \phi$, with uncorrelated white
 Gaussian thermal noise $\xi(t)$, 
$\langle \xi(t')\xi(t)\rangle =2 k_B T\eta\delta(|t-t'|)$. By introducing stochastic
dissipative force $F_d(t)=\eta\dot \phi-\xi(t)$, it can be understood as a force balance 
equation.
The net heat exchange with the environment is $Q(t)=\int_0^t \langle F_d (t')\dot \phi(t')\rangle dt'$ 
\cite{Sekimoto},
where $\langle...\rangle$ denotes ensemble average over a bunch of trajectory realizations.
Furthermore, $E_{\rm in}(t)=F\int_0^t \langle \dot \phi(t')\rangle dt'=
F \langle \phi(t)-\phi(0)\rangle$ is energy pumped into the
motor turnovers, and $W(t)=f_L\int_0^t \langle  \dot \phi(t')\rangle dt'=
f_L \langle \phi(t)-\phi(0)\rangle$ is the useful work
done against external torque. The fluctuations of motor energy 
$\int_0^t \langle f (\phi(t'))\dot \phi(t')\rangle dt'=
\int_0^{\phi(t)} \langle f (\phi)d\phi\rangle $ are bounded and can be neglected in the balance
of energies in a long run, since $Q(t)$, $E_{\rm in}(t)$, and $W(t)$ grow typically 
linearly, or possible also sublinearly (in case of anomalously slow dynamics with memory, see below),  in time. 
The energy balance yields the first law of 
thermodynamics: $Q(t)+W(t)=E_{\rm in}(t)$. The thermodynamic efficiency
is obviously 
\begin{eqnarray}
R=\lim_{t\to \infty}\frac{W(t)}{E_{\rm in}(t)}=\frac{f_L}{F}
\end{eqnarray}
independently of the presence of potential $V(\phi)$. It reaches unity at the stalling force
$f_L^{(st)}=F$. However, then the motor operates infinitesimally slow, $\dot \phi=\omega\to 0$.
Henceforth, a major interest present the efficiency $R_{\rm max}$ at the maximum of 
motor power $P_W=\dot W(t)$.
This one is easy to find in the absence of potential $V(\phi)$, i.e. for $f(x)=0$. Indeed,
$P_W=f_L\langle \dot \phi(t)\rangle=f_L(F-f_L)/\eta$. It shows a parabolic dependence
on $f_L$ and reaches the maximum at $f_L=F/2$. Therefore, $R_{\rm max}=1/2$.
Given this simple result many people believe until now that this is a theoretical 
bound for the
efficiency of isothermal motors at maximum power.

\subsubsection{Digression on the role of quantum fluctuations}

Within the simplest model considered ($f(x)=0$) the quantum noise effects do not play asymptotically 
any role for $T>0$.
This is not generally so, especially within a nonlinear dynamics, and at low temperatures, where
it can be dominant \cite{GoychukPRL98}. 
Most strikingly the role of zero-point fluctuations of vacuum (quantum noise at $T=0$) is demonstrated
in the Casimir effect: Two metallic plates will attract each other trying to minimize 
the ``dark energy'' of electromagnetic standing waves (quantized) in the space between 
two plates \cite{Lamoreaux}.
This effect can be used, in principle, to make a one-shot motor, which extracts energy from zero-point fluctuations
of vacuum, or ``dark energy'' doing work against an external force $f_L$. 
No violation of the second law of thermodynamics and/or the
law of energy conservation occurs, because such a ``motor'' cannot work cyclically.
In order to repeatedly extract energy from vacuum fluctuations, one must again separate two plates
and invest at least the same amount of energy in this. This example shows, nevertheless,
that the role of quantum noise effects can be highly nontrivial, very important, poorly understood, 
and possibly confusing. And a possibility to utilize ``dark energy'' to do a useful work
in a giant cosmic ``one-shot engine'' is really intriguing!

\subsection{Thermodynamic efficiency of isothermal engines
at maximum power can be larger than one half}

Let us show now that the belief that  $R_{\rm max}=1/2$ is a theoretical maximum is completely
wrong, and in accord with some recent studies
\cite{Schmiedl08,Seifert11,Broeck12,Golubeva} 
$R_{\rm max}$ can also achieve unity
within a nonlinear dynamics regime. For this, we first find stationary $\omega=\dot \phi$
in a biased periodic potential. This can be done by solving the Smoluchowski
equation for the probability density $P(x,t)$ which can be written as a continuity
equation $\partial P(\phi,t)/\partial t=-\partial J(\phi,t)/\partial \phi$, with 
the probability flux $J(x,t)$ written in the transport form
\begin{eqnarray}\label{fluxStrat}
J(\phi,t)=-De^{-\beta U(\phi)}\frac{\partial}{\partial \phi}e^{\beta U(\phi)}P(\phi,t).
\end{eqnarray}
This Smoluchowski equation  is an ensemble description counter-part
 of the Langevin equation (\ref{LE2}). Here, $D$  
is diffusion coefficient related to temperature and viscous friction by the
Einstein relation, $D=k_BT/\eta$, and $\beta=1/k_BT$ is inverse temperature. For any periodic
biased potential, the constant flux $J=\omega/(2\pi)= const$ driven by 
$\Delta \mu<0$,
as well as the corresponding non-equilibrium steady state distribution $P_{\rm st}(\phi)$
can be found by twice-integrating Eq. (\ref{fluxStrat}), using $U(\phi+2\pi)=U(\phi)-2\pi(F-f_L)$
 and periodicity of $P_{\rm st}(\phi)$.
This yields famous Stratonovich result \cite{Stratonovich58,StratonovichBook,RiskenBook} for a 
steady state angular velocity of phase rotation
\begin{eqnarray}\label{Strat}
\omega(\Delta \mu,f_L)=\omega_f(\Delta \mu,f_L)
\left [1- \exp(\beta (\Delta \mu+2\pi f_L)\right ]\equiv 
\omega_f(\Delta \mu,f_L)-\omega_b(\Delta \mu,f_L)
\end{eqnarray}
with forward rotation rate
\begin{eqnarray}\label{forward}
\omega_f(\Delta \mu,f_L)=\frac{2\pi D}{\int_{0}^{2\pi}d\phi \int_{\phi}^{\phi+2\pi}
e^{-\beta[U(\phi)-U(\phi')]}d\phi'}
\end{eqnarray}
and backward rate $\omega_f(\Delta \mu,f_L)$ defined by the
second equality in (\ref{Strat}). This result is quite general. 
$P_W(f_L)=f_L\omega_f(\Delta \mu,f_L)[1- \exp(\beta (\Delta \mu+2\pi f_L))$
and in order to find $R_{\rm max}$ one must find $f_L^{\rm (\max)}$ solving $ d P_W(f_L)/df_L=0$.
Then, $R_{\rm max}=f_L^{\rm (\max)}/F$. In fact, the relation (\ref{Strat})
is very general. It holds beyond the model of washboard potential leading
to result in (\ref{forward}). For example, given well defined potential minima
one can introduced picture of discrete states with classical Kramers rate for
the transitions between those, see in Fig. \ref{Fig1}, (b). Accordingly, 
within  a simplest enzyme model, one has
three discrete states. $\textbf{E}$ corresponds to empty enzyme with energy
$E_1$.  $\textbf{ES}$ corresponds to enzyme with substrate molecule bound and energy
$E_2$ of the whole complex.  $\textbf{EP}$ corresponds to enzyme with product
 molecule(s) bound and energy $E_3$. The forward cyclic transitions 
 $\textbf{E}\to \textbf{ES}\to \textbf{EP}\to \textbf{E}$ are driven by free energy per one
 molecule $\Delta \mu$ released in $\textbf{S}\to \textbf{P}$ transformation facilitated
 by enzyme, while the backward cycling $\textbf{E}\to \textbf{EP}\to \textbf{ES}\to 
 \textbf{E}$ requires backward reaction $\textbf{P}\to \textbf{S}$. It is normally neglected
 in the standard Michaelis-Menthen type approaches to enzyme kinetics as one very unlikely to occur.
 This generally cannot be done for molecular motors.
 The simplest possible Arrhenius 
 model for forward rate of the whole cycle reads
 \begin{eqnarray}\label{Arrhen}
 \omega_f(\Delta \mu,f_L)=\omega_0\exp[-\beta\delta (\Delta \mu+2\pi f_L)]
 \end{eqnarray}
where $0<\delta<1$ describes asymmetry of potential drop.
The backward rate is accordingly, 
$ \omega_b(\Delta \mu,f_L)=\omega_0\exp[\beta(1-\delta) (\Delta \mu+2\pi f_L)]$.
This model allows to realize under which conditions $R_{\rm max}$ can exceed one-half. 
Here we rephrase
a recent treatment in \cite{Schmiedl08,Seifert11} and come to the same conclusions.
$R_{\rm max}$ is solution of $dP_W(f_L)/df_L=0$, which leads to a transcendental 
equation for $R_{\rm max}$
\begin{eqnarray}\label{eq_for_Rmax1}
\exp [r (1-R_{\rm max})]=1+\frac{r R_{\rm max}}{1+r b R_{\rm max}},
\end{eqnarray}
where $r=|\Delta \mu|/(k_B T)$, and 
$b=(k_BT/2\pi)\partial \ln \omega_f(\Delta \mu,f_L)/\partial f_L$. For (\ref{Arrhen}),
$b=-\delta$. The limiting case $b=0$ of extreme
asymmetry is especially insightful. In this special case,
$R_{\rm max}=[LW(e^{1+r})-1]/r$ exactly,
where $LW(z)$ denotes the Lambert W-function. This analytical
result shows that $R_{\rm max}\to 1/2$ as $r\to 0$,
while $R_{\rm max}\to 1$ as $r\to \infty$. Therefore, a popular statement
that $R_{\rm max}$ is generally bounded by $1/2$ is simply wrong. 
Yes, in some models this Jacobi bound exists, but generally not. Already the simplest
model of molecular motors considered here following \cite{Schmiedl08} completely refutes 
the Jacobi bound as a theoretical limit.
Further insight emerges in the perturbative regime, $r\ll 1$, which yields
in the lowest order of $r$
\begin{eqnarray}\label{perturb1}
R_{\rm max}&= &\frac{1}{2}+\frac{1}{8}\left (\frac{b+1}{2} \right)r + o(r)\nonumber \\
&\approx &\frac{1}{2}+\frac{1}{8}\left (\frac{1}{2}-\delta \right)\frac{|\Delta \mu|}{k_BT}\;.
\end{eqnarray}
This is essentially the same result as in \cite{Seifert11}. Hence,
for $0\leq \delta <1/2$, $R_{\rm max}>1/2$ for a small $r$. The effect is small for $r\ll 1$, but
it exists.

\begin{figure}
\includegraphics[width=16cm,keepaspectratio]{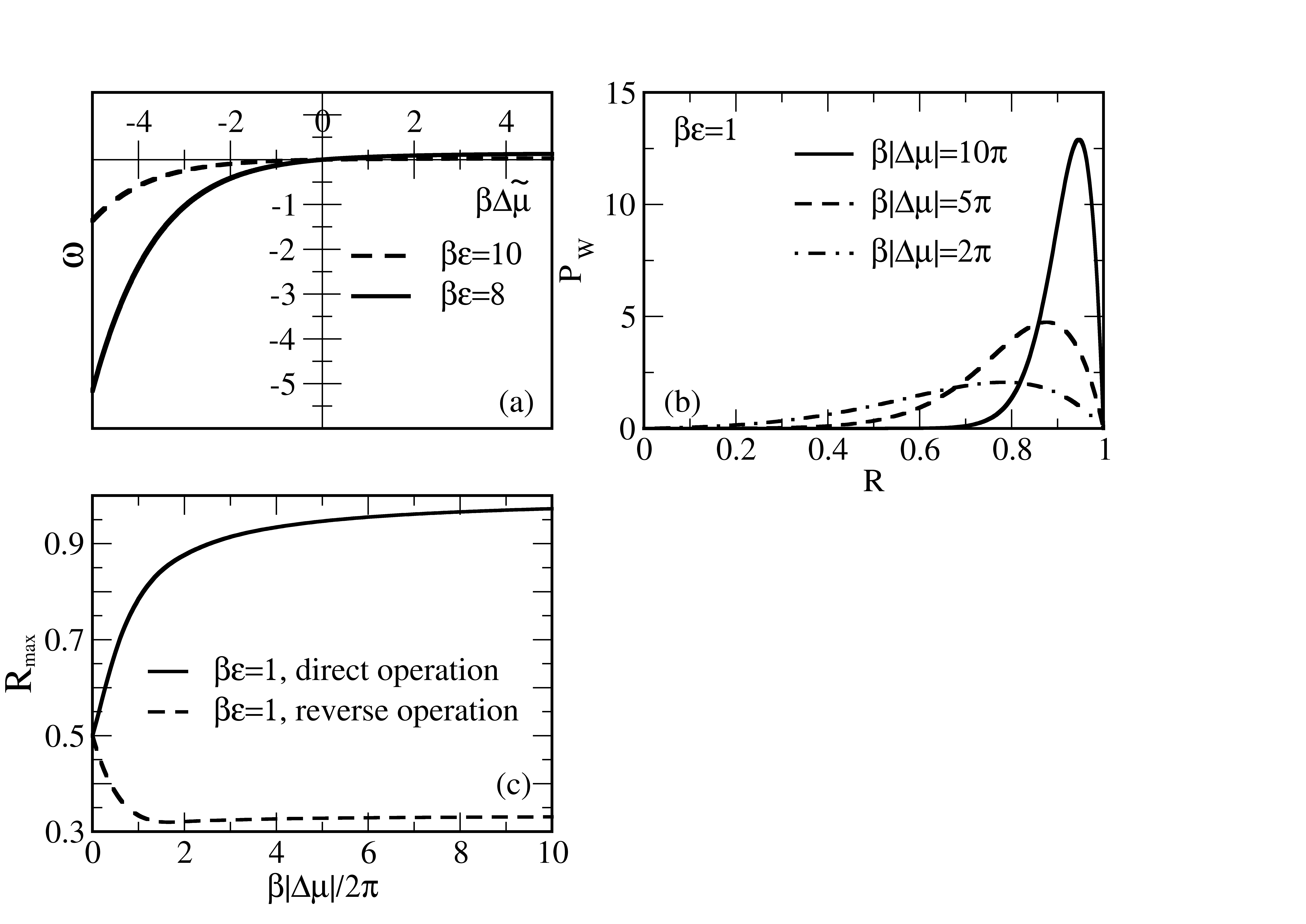}
\caption{(a) Dependence of the net rotation rate $\omega$ on
the net bias $\widetilde{ \Delta \mu}$ for the most asymmetric 
sawtooth model depicted in Fig. \ref{Fig1} (a), for two values of the effective
barrier height $\beta\epsilon$. (b) Dependence of the output power $P_W$ 
on thermodynamic efficiency
$R$ for $\beta\epsilon=1$ and several values of the scaled driving force
$\beta |\Delta \mu|$. (c) The maximum power efficiency 
as function of driving force for the direct and inverse operation, when
the roles of driving force and load are interchanged. }
\label{Fig2} 
\end{figure}

The discussed model  might seem a bit too crude. However, the result that $R_{\rm max}$
can achieve theoretical limit of unity survives also within a more advanced, 
yet very simple model. Indeed, 
let us consider the simplest
kind of sawtooth potential in Fig. \ref{Fig1} inspired by the above discrete
state model with $\delta=0$. Then, Eq. (\ref{forward}) yields explicitly,
\begin{eqnarray}\label{sawtooth}
\omega_f(\Delta \mu,f_L)=\frac{ D}{8\pi} 
\frac{\beta^2 \epsilon (\epsilon - \Delta \mu-2\pi f_L)}
{\sinh^2(\beta \epsilon/2)-[\epsilon/( \Delta \mu+2\pi f_L)]
\sinh^2(\beta ( \Delta \mu+2\pi f_L)/2)}\,.
\end{eqnarray}
The dependence of $\omega(\Delta \mu,f_L)$ on $\widetilde{ \Delta \mu}:=
|\Delta \mu|-2\pi f_L$ is very asymmetric within this model, see in Fig. \ref{Fig2}, (a).
\vspace{0.5cm}
This is a typical diode type or rectifier dependence, if to apply the same model 
to transport of charged particles in a spatially periodic potential, with 
$\omega(\Delta \mu,f_L)$ corresponding to a scaled current and $\widetilde{ \Delta \mu}$
to voltage. Clearly, within the latter context, if to apply an additional sufficiently 
slow periodic
voltage signal $A\cos(\omega t)$ at the conditions $\widetilde{ \Delta \mu}=0$, it will
be rectified because of asymmetric $I-V$ characteristics, giving rise to a directional
dissipative current in
a potential unbiased on average (both spatial and time averages are zero). 
The effect gave rise to a huge literature
on rocking Brownian ratchets, in particular, and on Brownian motors, in general,
see e.g. \cite{ReimannReview} for a review. 
Turning back to the efficiency
of molecular motor  at maximum power within our model, we see clearly in Fig. \ref{Fig2}, (c)
that it can be well above $1/2$, and even close to one. 
A sharply asymmetric dependence of $P_W$
on $R=f_L/F$ in Fig. \ref{Fig2}, (b) beyond linear response regime 
$P_W=4P_{\rm max}R(1-R)$, which is not
shown therein because of a very small $P_{\rm max}$, provides an additional
clue on the origin of this remarkable effect. Interestingly, if to reverse the work
of our motor, i.e. $f_L $ provides supply of energy and a useful work is done against
$F\leq f_L$,  then the motor rotates in the opposite direction on average. 
This occurs, for example, in such enzymes as F0F1-ATPase \cite{Pollard08,Nelson,Yoshida},
which presents a complex of two rotary motors  F0 and F1 connected by a common shaft.
F0 motor uses electrochemical gradient of protons to rotate the shaft which transmits
the torque on F1 motor. The mechanical torque
applied to F1 motor is used to synthesize ATP out of ADP and phosphate group $\rm P_i$.
This enzyme complex utilizes primarily electrochemical gradient of protons to synthesize ATP.
It can, however, also work in reverse
and pump protons using energy of ATP hydrolysis \cite{Yoshida}. Moreover,  in a separated
F1-ATPase motor, the energy of ATP hydrolysis can be used to create a mechanical torque
and do a useful work against external load, which is well studied experimentally \cite{Toyabe}.
For the reverse operation, our minimalist motor 
efficiency becomes $R'=P'_W/P'_{\rm in}$, where $P'_W=F |\langle \dot \phi(t)\rangle|$, 
and $P'_{\rm in}=f_L |\langle \dot \phi(t)\rangle|$.
In this case, $R'_{\rm max}$ indeed cannot exceed $1/2$, see in Fig. \ref{Fig2}, (c),
the lower curve. 
Such a behavior is also to expect from the above discrete state model, because this corresponds
to $\delta\to 1=-b$ in (\ref{perturb1}). This argumentation can be inverted:
If a motor obeys the Jacobi bound, $R_{\rm max}\leq 1/2$, then working in reverse it
can violate it. Hence, understanding of Jacobi bound as a fundamental one 
is clearly a dangerous misconception which should be avoided.

\subsection{Minimalist model of quantum engine}

\begin{figure}
\includegraphics[width=12cm,keepaspectratio]{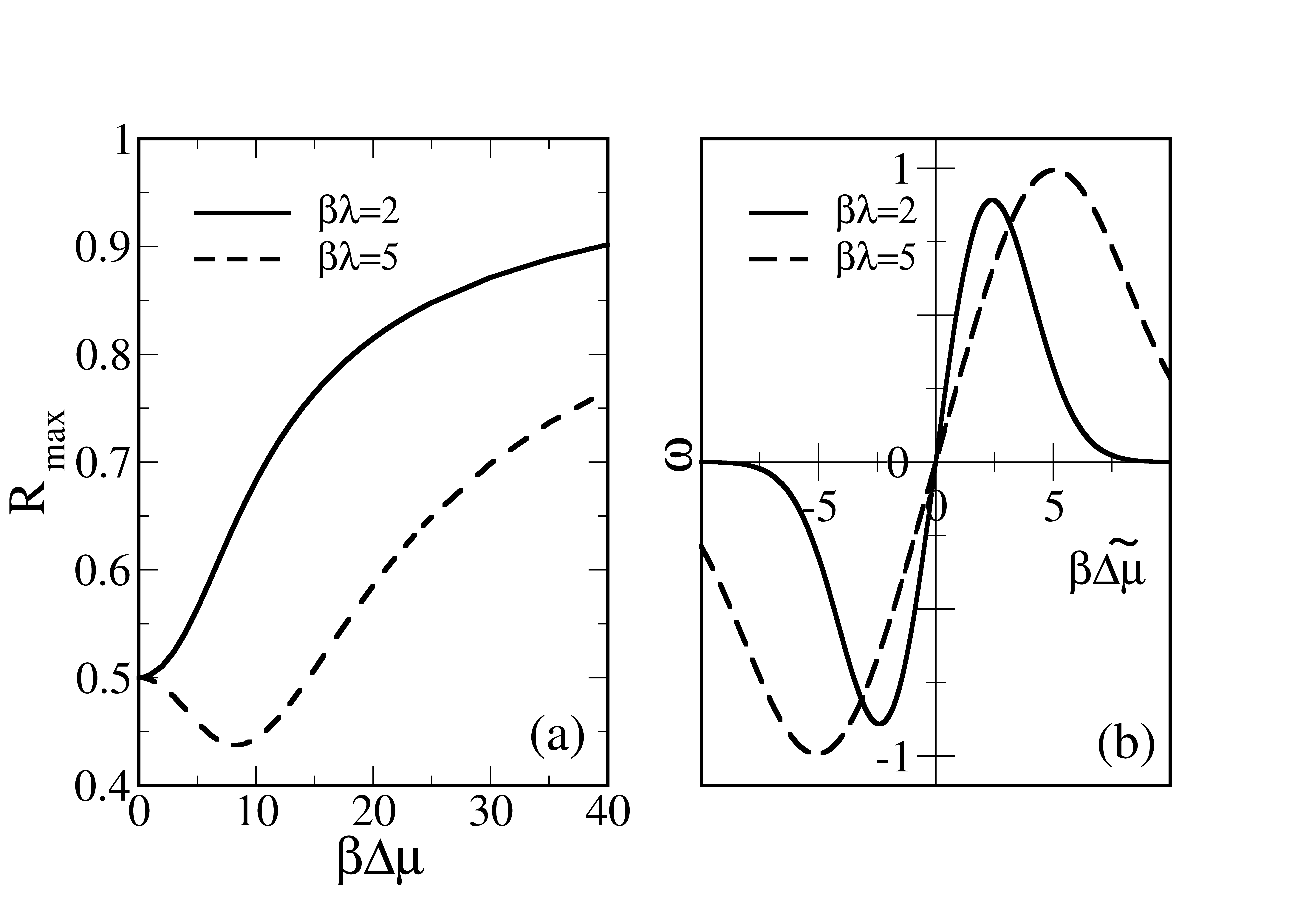}
\caption{(a) Dependence of $R_{\rm max}$ on the absolute value of 
driving energy $\Delta \mu$ in units of $k_BT$
for two values of $\lambda/k_BT$.
Within the perturbative regime Eq. (\ref{perturb2}) predicts well the initial dependence. 
(b) Dependence of enzyme velocity
$\omega$ on $\beta\widetilde{\Delta \mu}$. Notice the existence of maximal $\omega$ and negative
differential regime.}
\label{Fig3} 
\end{figure}

In quantum case, discrete state models emerge naturally. For example, energy levels depicted
in Fig. \ref{Fig1}, (b) can correspond to the states of a proton pump driven by a nonequilibrium 
electron flow. This is a minimalist toy model for pumps like
cytochrome c oxidase proton pump \cite{Pollard08,Wikstroem}. 
The driving force is provided by electron energy
 $\Delta \mu$ released by dissipative tunneling of electron between donor and acceptor electronic states
 of the pump. This process is complex. It requires, apart from intramolecular electron transfer,
also uptake and release of electrons from two bath of electrons on different
sides of membrane, which can be provided e.g. by mobile electron carriers \cite{Pollard08}. 
However, intramolecular electron transfer (ET) between two heme metalloclusters 
seems to be a rate limiting step. Such ET presents
vibrationally-assisted electron tunneling between two localized quantum states 
\cite{Atkins,Nitzan}. 
Given weak electron
tunnel coupling between electronic states, the rate can be calculated e.g. using 
quantum-mechanical Golden Rule.
Within classical approximation of nuclei dynamics (but not one of electrons!), and simplest possible
further approximations one obtains the celebrated Marcus-Levich-Dogonadze rate
\begin{eqnarray}\label{MLD}
\omega_f(\Delta \mu,\Delta \mu_p=0)=\omega_0\exp\left 
[-(\Delta \mu+\lambda)^2/(4\lambda k_BT) \right ]
\end{eqnarray}
for forward transfer, and $\omega_b(\Delta \mu,\Delta \mu_p=0)=
\omega_f(\Delta \mu,\Delta \mu_p=0)\exp[\Delta \mu/(k_BT)]$. Here, 
$\omega_0=(2\pi/\hbar)V_{\rm tun}^2/\sqrt{4\pi \lambda k_BT}$ is a
quantum prefactor, where $V_{\rm tun}$ is tunnel coupling, and $\lambda$ is medium's
reorganization energy. Energy released in the electron transport is used 
to pump protons against their electrochemical gradient $\Delta \mu_p$, which 
corresponds to $2\pi f_L$  within the previous model. Hence, $R=\Delta \mu_p/|\Delta \mu|$.
Of course, our model should not 
be considered as a realistic model for cytochrome c oxidase. However, it allows
to highlight a possible role of quantum effects which are contained in the dependence
of the Marcus-Levich-Dogonadze rates on the energy bias $\Delta \mu$. Namely, the existence of
 inverted ET regime  when the rate becomes smaller with a further increase of 
$|\Delta \mu|>\lambda$, after reaching a maximum at $|\Delta \mu|=\lambda$ (activationless
regime). The inverted regime is a purely quantum-mechanical feature. It cannot
be realized within a classical adiabatic Marcus-Hush regime, for which the 
rate expression is looking formally the same as (\ref{MLD}), however, with a classical
prefactor $\omega_0$. Classically, inverted regime makes  simply no literal physical sense. 
This fact can be easily realized upon plotting the lower adiabatic curve
for underlying curve crossing problem (within the Born-Oppenheimer approximation), 
and considering the pertinent activation barriers -- the way how 
Marcus parabolic dependence of the activation energy on the energy 
bias is derived in textbooks \cite{Atkins}. 
The fact that inverted ET regime
can be used to pump electrons has been first realized within a driven spin-boson
model \cite{GoychukCPL96,GoychukPRE97,GoychukJCP97,GoychukAP05}. The model here is, however, 
very different, and pumping
 is not relied on inverted ET regime. However, the latter one can be used to arrive
at a high $R_{\rm max}$ close to one. Indeed, within this model the former (Arrhenius rates)
parameter $b$ becomes $b=-1/2+(|\Delta \mu|-\Delta \mu_p)/(4\lambda)$, 
and Eq. (\ref{eq_for_Rmax1}) is replaced now by 
\begin{eqnarray}\label{eq_for_Rmax2}
\exp [ r (1-R_{\rm max})]=\frac{1+
r \left [ -1/2+ r(1-R_{\rm max})/(4c) \right ]R_{\rm max}}{1+
r \left [ 1/2+ r(1-R_{\rm max})/(4c) \right ]R_{\rm max}}\;.
\end{eqnarray}
A new control parameter $c=\lambda/(k_BT)$ enters this expression. 
Perturbative solution of (\ref{eq_for_Rmax2}) for $r=|\Delta \mu|/k_BT\ll 1$ yields
\begin{eqnarray}\label{perturb2}
R_{\rm max}\approx \frac{1}{2}+\frac{1}{192}\frac{(\Delta \mu)^2}{\lambda k_BT }
\left (3-\frac{\lambda}{k_BT} \right )
\end{eqnarray}
to the lowest second order in $|\Delta \mu|/k_BT$ (compare with (\ref{perturb1})!). 
Hence, $R_{\rm max}>1/2$  for $\lambda<3 k_BT$
and $R_{\rm max}<1/2$  for $\lambda>3 k_BT$
in the perturbative regime.
However, beyond it  $R_{\rm max}$  can essentially be
larger than $1/2$, see in Fig. \ref{Fig3}, (a).

These results are also to
expect, for the pump working in reverse when $\Delta \mu\to -\Delta \mu$. 
Here, we also see a huge difference with the model based on Arrhenius rates. 
The dependence of 
rotation rate $\omega$ on
$\widetilde{\Delta \mu}=|\Delta \mu|-\Delta \mu_p$ in this case is symmetric.
However, it exhibits a regime with negative differential part, where 
$d\omega /d\widetilde{\Delta \mu}<0$, for $\widetilde{\Delta \mu}$ exceeding
some critical value, which approaches $\lambda$ for small $T$, see
in Fig. \ref{Fig3}, (b). Here, the reason for a high performance is very different
from the case of asymmetric Arrhenius rates, or asymmetric $V(\phi)$.
$R_{\rm max}$ can be close to one for $|\widetilde{\Delta \mu}|\gg \lambda$. For this to happen, 
the motor should be driven deeply into the inverted ET regime. 
Hence, the effect is quantum-mechanical in nature, even if the considered setup looks
purely classical. In this respect, Pauli quantum master equation for the diagonal
elements of reduced density matrix decoupled from the off-diagonal
elements has mathematical form of
classical master equation for population probabilities, and the corresponding
classical probability description can be safely used. The rates entering this
equation can, however, reflect such profound quantum effects as quantum-mechanical tunneling
and yield non-Arrhenius dependencies of dissipative tunneling rates on temperature and
external forces. The corresponding quantum generalizations of classical 
results become rather straightforward.
The theory
of quantum nanomachines with profound quantum coherence effects 
is, however, still in infancy. \\ \vspace{1cm}

\subsection{Can rocking ratchet do a useful work without dissipation?}
 
 As we just showed, strong dissipation is not an obstacle for either classical, or quantum 
 Brownian machines to achieve a theoretical limit of performance. This already indicates
 that to completely avoid dissipation is neither possible, nor even desirable
 to achieve in nanoworld to build a good nanomachine. 
 Vice versa, the so-called rocking ratchets without dissipation \cite{Flach00,Goychuk00} are
 not capable to do any useful work, in spite of they can produce a directional
 transport. However, this directional transport not necessarily  can proceed against any non-zero 
 force trying to stop it, as we will now proceed to clarify. The stalling force can become negligibly small, 
 and thermodynamical efficiency of such a device is zero, very differently from genuine ratchets,
 which must be characterized by a non-zero stalling force \cite{ReimannReview}.
 Therefore, ratchet current without dissipation presents clearly
 an interesting but futile artefact. The rocking ratchets without dissipation
 should be named pseudo-ratchets to distinguish them from genuine ratchets characterized 
 by non-zero stalling force. 
 
 Let us consider the following setup. The particle in a periodic potential $V(x)$ is
 driven by a time-periodic force $g(t+{\cal T})=g(t)$, with period ${\cal T}$. Then, 
 $U(x,t)=V(x)-xg(t)$, or $f(x,t)=f(x)+g(t)$ in Eq. (\ref{LE}). For strong dissipation and 
 overdamped Langevin dynamics, $M\to 0$, the rectification current can emerge in potentials
 with broken space-inversion symmetry, like one in Fig. \ref{Fig1}, (a) under a 
 fully symmetric driving like $g(t)=A\cos(\Omega t)$,
 $\Omega =2\pi/{\cal T}$. Broken space-inversion symmetry means that is there is no
 such $x_0$, that $V(-x)=V(x+x_0)$. Likewise, a periodic driving is symmetric with respect to
 time reversal, if there exist such a $t_0$ (or equivalently, a phase shift 
 $2\pi t_0/{\cal T}$), such that $g(-t)=g(t+t_0)$, and breaks the time-reversal symmetry
 otherwise. Also, higher moments of driving, $\overline{g^n(t)}=
 \frac{1}{{\cal T}}\int_0^{\cal T} g^n(t)dt$, $n=2,3,...$
 are important with respect to a nonlinear response reasoning. The latter moments
 can be defined also for stochastic driving, using a corresponding 
 time-averaging, with $\cal T\to\infty$. For overdamped dynamics,
 rectification current appears already in the lowest second order of 
 $\overline{g^2(t)}\neq 0$, for a potential with broken spatial-inversion symmetry, 
 and in the  the lowest third order of $\overline{g^3(t)}\neq 0$ for potentials which are
 symmetric with respect to inversion $x\to -x$ \cite{ReimannReview}. These results were
 easy to anticipate for memoryless dynamics, which displays asymmetric current-force
 characteristics in the case of static force applied (broken spatial symmetry), or a symmetric
 one (unbroken symmetry), respectively. They hold also quantum-mechanically
 in the limit of strong dissipation. The case of weak dissipation is, however, more intricate
 both classically and quantum-mechanically. A symmetry analysis based on Curie symmetry principle
 has been developed in order to clarify the issue \cite{Flach00,ReimannReview}. 
 The harmonic mixing driving \cite{Wonneberger},
 \begin{eqnarray}\label{HM}
 g(t)=A_1\cos(\Omega t+\phi_0)+A_2\cos(2\Omega t+2\phi_0+\psi),
 \end{eqnarray}
 is especially interesting in this respect. 
 Here, $\psi$ is a relative phase of two harmonics, which plays a crucial role, and $\phi_0$
 is an absolute initial phase, which physically cannot play any role because it corresponds
 to a time shift $t\to t+t_0$ with $t_0=\phi_0/\Omega$ and hence must 
 be averaged out in final results, if they are of any physical importance in real world. 
 Harmonic
 mixing driving provides a nice testbed, because this is a simplest time-periodic driving
 which can violate the time-reversal symmetry, which happens for any $\psi\neq 0,\pi$. On the other
 hand,  $\overline{g^3(t)}=(3/4)A_1^2A_2\cos(\psi)$. Hence, $\overline{g^3(t)}\neq 0$, for
 $\psi\neq \pi/2,3\pi/2$. Interestingly, $\overline{g^3(t)}=(3/4)A_1^2A_2$ is 
 maximal for time-reversal
 symmetric  driving, and vice versa $\overline{g^3(t)}=0$, when the time reversal symmetry
 is maximally broken. Moreover, one can show that all odd moments 
  $\overline{g^{2n+1}(t)}\propto \cos(\psi)$, $n=1,2,3,...$, vanish for
  $\psi=\pi/2$ or $3\pi/2$. Vanishing of odd moments for a periodic function
  means that it obeys a symmetry condition $g(t+{\cal T}/2)=-g(t)$.
  Also in application to potentials of the form $V(x)=V_1\cos(kx)+V_2\cos(2kx+\phi)$, these
  results mean that $\overline{V^3(x)}\propto \cos(\phi)$, and  
  $\overline{f^3(x)}\propto \sin(\phi)$, for the corresponding spatial averages. Hence,
  for a space-inversion symmetric potential with $\phi=0$, $\overline{f^3(x)}=0$ (also
  all higher odd moments vanish). Moreover, $\overline{f^3(x)}$ is maximal, when the
  latter symmetry is maximally broken, $\phi=\pi/2$. This corresponds to the so-called
  ratchet potentials. The origin of rectification current can be understood as a memoryless
  nonlinear response in the overdamped systems: For $\overline{f^3(x)}\neq 0$, the current
  emerges already for standard harmonic driving, as a second order response to driving.
  For $\overline{f^3(x)}=0$ (e.g. standard cosine potential, $V_2=0$), one needs 
  $\overline{g^3(t)}\neq 0$ for driving to produce the ratchet effect. For the above harmonic driving, the 
  averaged current $\langle \dot x(t)\rangle \propto \cos(\psi)$. The same type of
  response behavior features also quantum-mechanical dissipative single-band tight-binding
  model for strong dissipation \cite{GoychukEPL98,GoychukJPCB}. Very important is that any genuine fluctuating tilt 
  or rocking ratchet is characterized
  by a non-zero stalling force, which means that  the ratchet transport can sustain
  against a loading force and do useful work against it. It ceases at a critical stalling force.
  This has important implications.
  For example, in application to photovoltaic effect in crystals with broken space-inversion
  symmetry \cite{ReimannReview} this means that two opposite surfaces of crystal (orthogonal to current flow) 
  will be gradually charged 
  until the emerged photo-voltage will stop the ratchet current flow. For zero stalling force,
  no steady-state photo-voltage or electromotive force can in principle emerge!  
  
\begin{figure}
\includegraphics[width=12cm,keepaspectratio]{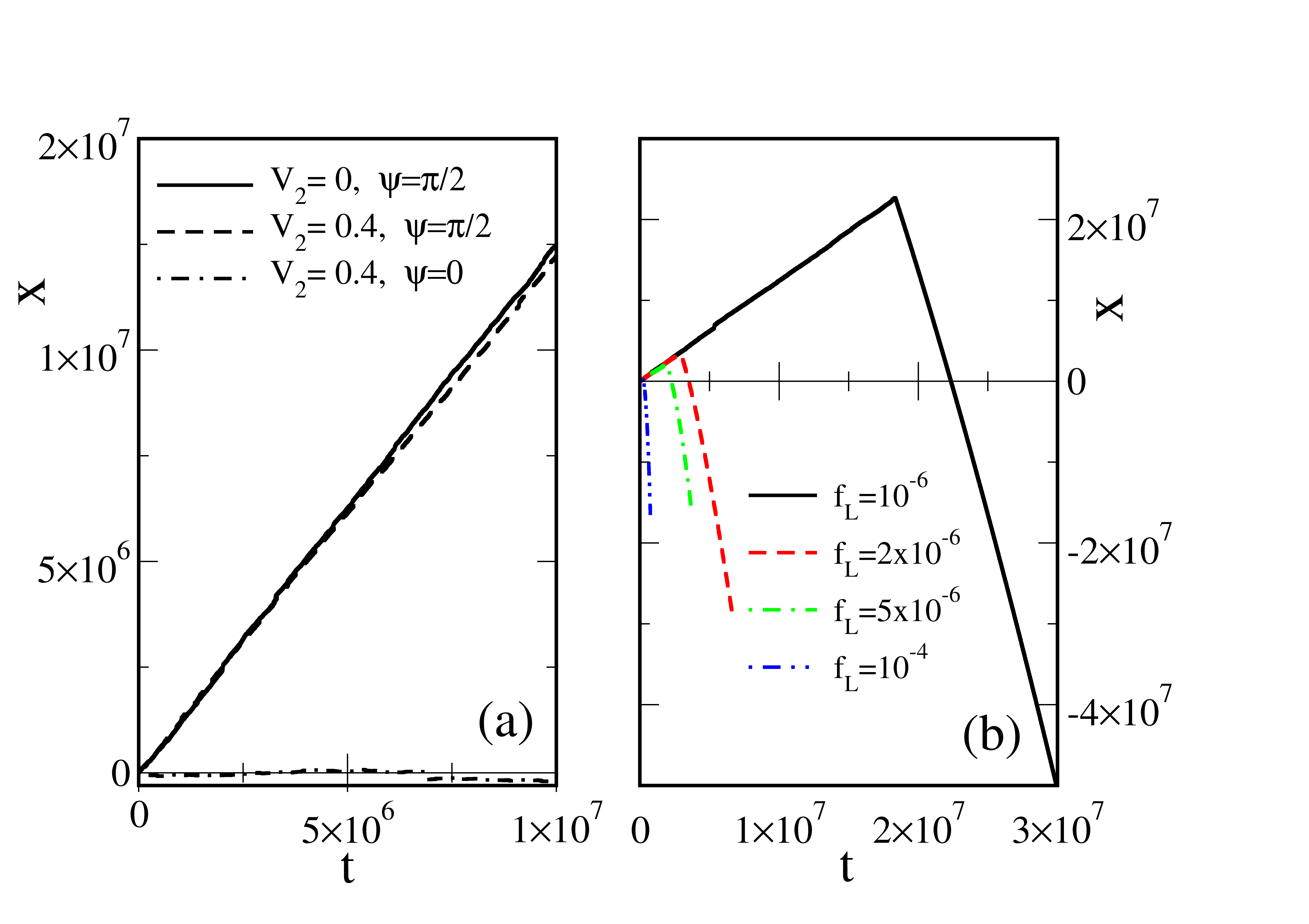} 
\caption{(a) Directed transport for standard cosine potential, $V_1=1$, $V_2=0$, and for a ratchet
potential,  $V_1=1$, $V_2=0.4$, in the case of harmonic mixing driving that breaks the time-reversal
symmetry, $\psi=\pi/2$, with amplitudes $A_1=5$, $A_2=2$, and frequency $\Omega=1$.
Transport ceases at $\psi=0$ even for ratchet potential with broken symmetry, when the time-reversal
symmetry is restored (dash-dotted line). (b) Influence of a tiny (as compare with the periodic force modulation)
constant loading force on transport for the case of ratchet potential in part (a). The transport ceases
after some random time, which depends on  $f_L$ and initial conditions, and the particle returns
accelerating  back. Very similar picture emerges also for cosine potential with $V_2=0$ (not shown).
The stalling force is obviously zero. Any genuine ratchet and motor must be characterized by
a non-zero stalling force. Symplectic leapfrog/Verlet integration scheme (there is no spurious
dissipation introduced by numerics) was used to obtain these results.
}
\label{Fig4}
\end{figure}

  In the case of weak dissipation, however, memory effects
  in the current response become essential. Generally, for classical dynamics, 
  $\langle \dot x(t)\rangle \propto \cos(\psi-\psi_0)$, where $\psi_0$ is a phase shift
  which depends on the strength of dissipation with two limiting cases: (i) $\psi_0=0$ for for overdamped
  dynamics, and (ii) $\psi_0\to \pi/2$ for vanishing dissipation $\eta\to 0$. In the later limit, the
  system becomes purely dynamical:
  \begin{equation}\label{DE}
M\ddot x=f(x)+g(t)-f_L\;,
\end{equation}
where we added an opposing transport loading force $f_L$. For example, it corresponds
to a counter-directed electrical field in the case of charged particles.
Let us consider following \cite{Flach00,Goychuk00}, the two original papers on 
dissipationless ratchet current in the case of $f_L=0$, the following potential
$V(x)=-V_1 \sin(2\pi x)-V_2\sin(4\pi x)$, or $f(x)=f_1\cos(2\pi x)+f_2\cos(4\pi x)$,
with $f_1=2\pi V_1$, $f_2=4\pi V_2$, and driven by $g(t)$ in (\ref{HM}). 
The spatial period is one, and $M=1$ in dimensionless units.
The emergence of dissipationless current within the considered dynamics 
has been rationalized within a symmetry analysis
 in \cite{Flach00}, and the subject of directed currents due to broken time-space 
 symmetries has been born. In an immediate follow-up work \cite{Goychuk00}, we have, however,
 observed that in the above case, the directed current is produced only by breaking the 
 time-reversal symmetry by a time-dependent driving, and not otherwise. 
 Breaking of spatial symmetry of the potential alone does not originate dissipation-less current.
 The current is maximal at $\psi=\pi/2$.
  No current emerges, however, at $\psi=0$ even in ratchet potential with broken space-inversion
  symmetry. Moreover, the presence of second potential harmonic does not seem to affect 
  the transport at  $\psi=\pi/2$, see in Fig. \ref{Fig4}, (a) two cases which differ by
  $V_2=0$, in one case, and $V_2\neq 0$, in another one. 

  Moreover, within a corresponding
  Langevin dynamics, when dissipation is present,
   each and every trajectory remains time-reversal symmetric for $\psi=0$. However, for
  a strongly overdamped dynamics the rectification current in a symmetric cosine potential
  ceases at $\psi=\pi/2$, and not at $\psi=0$. Moreover, for an intermediate dissipation
  it stops at some $\psi_0$, $0<\psi_0<\pi/2$, as in \cite{Dykman97}. Which symmetry forbids it then, 
  at a particular non-zero dissipation
  strength? Dynamical symmetry considerations fail to answer such simple questions, and
  are thus not almighty.
  The symmetry of individual trajectories  within a Langevin description simply does not depend
  on the dissipation strength, which can be easily understood from a well-known dynamical
  derivation of this equation we presented above.
  Therefore, a symmetry argumentation based on the symmetry properties of single trajectories is
  clearly questionable, in general. Spontaneous breaking of symmetry 
  is a well-known fundamental phenomenon both in quantum field theory and the theory of phase
  transitions. In this respect, any chaotic  Hamiltonian dynamics 
  possesses the following symmetry:
  for any positive Lyapunov exponent, there is a negative Lyapunov exponent having the same
  absolute value of real part. The time reversal changes the sign of Lyapunov exponents. This symmetry
  is spontaneously broken in Hamiltonian dynamics by considering forward 
  evolution in time \cite{Gaspard}. It becomes especially obvious
   upon making coarse-graining
  which is not possible to avoid neither in real life, nor in numerical experiments.
  By the same token, time-irreversibility of Langevin description given time-reversible trajectories
   is primarily statistical
  and not dynamical effect.
  
  The emergence of such a current without dissipation has been interpreted as a 
  reincarnation of Maxwell-Loschmidt demon \cite{Goychuk00}, and
  it has been argued that this demon is killed by a stochastically fluctuating \textit{absolute}
  phase $\phi_0\to \phi_0(t)$, with the relative phase $\psi$ \textit{being fixed}. In this respect, 
  even in highly coherent sources of light  such as lasers  the absolute phase fluctuations
  cannot be avoided in principle. They yield a finite bandwidth of laser light. 
  The phase shift $\psi$ can be stabilized,
  but not the absolute phase. Typical dephasing time of semiconductor lasers used in laser
  pointers is in nanoseconds range, whereas in long tube lasers it is improved to
  milliseconds  \cite{Paschotta}. This is the reason why some averaging over such fluctuations
  must always be done, see in \cite{RiskenBook}, Ch. 12.
  The validity of this argumentation has been analytically
  proven in \cite{Goychuk00} with an exactly solvable example of quantum-mechanical
  tight binding model driven by harmonic mixing drive with a dichotomously fluctuating $\phi_0(t)$.
  Even more spectacularly this is seen in a dissipationless tight-binding dynamics driven
  by an asymmetric stochastic two-state field. Current is completely absent even for 
  $\overline{g^{2n+1}(t)}\neq 0$, as an exact solution shows \cite{GoychukPRL98}.
  Hence, dissipation is required to produce a ratchet current under a \textit{stochastic} driving $g(t)$.
  The validity of this result is far beyond the particular models 
  in \cite{GoychukPRL98,GoychukEPL98,Goychuk00} because any \textit{coherent} quantum current,
  e.g. one carried by Bloch electron with non-zero quasi-momentum is killed by quantum decoherence
  produced by a stochastic field. Any dissipationless quantum current will proceed on the time scale
  smaller than decoherence time.
  
  Moreover, here we show that the directed transport without dissipation found in \cite{Flach00,Goychuk00},
  and follow-up research work cannot do any useful work against an opposing force $f_L$. Indeed,
  the numerical results shown in Fig. \ref{Fig4}, (b) reveal this clearly: After some random time
  (depends, in particular, on initial conditions and on the load $f_L$ strength), the rectification
  current ceases. As a matter of fact, the particle moves then back much faster, with acceleration. 
  The smaller $f_L$, the longer is the directional \textit{normal} 
  transport regime and smaller back acceleration, and nevertheless the forward transport 
  is absent asymptotically.
   Therefore, this ``Maxwell demon'' cannot do asymptotically any useful work, unlike e.g.
  highly efficient ionic pumps -- the ``Maxwell demons'' of living cells working under condition
  of strong friction.  
  Plainly said, dissipationless demon cannot charge
  a battery, it is futile. Therefore, to consider such a device as ``motor'' cannot 
  be scientifically justified. Clear is also that with vanishing friction thermodynamic 
  efficiency of rocking Brownian motors also vanishes. Therefore, a naive feeling 
  that smaller friction provides higher efficiency is completely wrong, in general.

 
Let us briefly summarize the major findings of this section. First, friction and noise are 
intimately related in 
microworld which is nicely seen from a mechanistic derivation of (generalized) Langevin dynamics.
It results from a hyper-dimensional Hamiltonian dynamics with random initial conditions like in
molecular dynamics approach. For this reason, thermodynamic efficiency of isothermal nanomotors
can arrive 100\% even under conditions of a very strong dissipation, in the overdamped regime
where the inertial effects become negligible. Quite on the contrary, thermodynamical efficiency
of low-dimensional dissipationless Hamiltonian ratchets is zero. Therefore, they cannot serve
as a model for nanomotors in condensed media. Moreover, some current realizations of Hamiltonian
ratchets with optical lattices exceed in geometrical sizes such nanomotors as F1-ATPase by
several orders of magnitude. In this respect, the readers should be reminded that a typical wavelength
of light is about a half of micron which is the reason why such motors as  F1-ATPase cannot be seen
in a standard light microscope. Hence, the whole subject of Hamiltonian dissipationless ratchets
is completely irrelevant for nanomachinery. 
Second, thermodynamical efficiency at maximum power in nonlinear 
regimes can well exceed the upper bound of 50\% valid only for a linear dynamics. Therefore, nonlinear
effects are generally very important to build up a highly efficient nanomachine. Third, important
quantum effects can be already captured within a rate dynamics with quantum rates obtained e.g.
using a quantum-mechanical perturbation theory in tunnel coupling, i.e. within a Fermi's Golden Rule
description whose particularly simple limit results into Marcus-Levich-Dogonadze rates of nonadiabatic
tunneling.



\section{Adiabatic pumping and beyond }

Having realized that thermodynamic efficiency at maximum power can exceed 50\%, a natural
question emerges: How to arrive at such an efficiency in practice?
Intuitively, the highest thermodynamical efficiency of molecular and other nanomotors
can be achieved for an adiabatic modulation of potential when the potential is gradually
deformed so that its deep minimum gradually moves from one place to another one and a
particle trapped near to this minimum follows adiabatic modulation of the 
potential in a peristaltic like motion.
The idea is that the relaxation processes are so fast (once again, a sufficiently  strong dissipation
is required!) that they occur almost instantly on the time scale of potential modulation.
In such a way, the particle can be transferred in a highly dissipative
environment from one place to another one practically without heat losses, and do a useful 
work against a substantial load, see e.g. discussion
in \cite{AstumianPNAS}.
If at any instance of time, the motor particle stays always near to thermodynamic equilibrium, 
then in accordance  with FDT the total heat losses to the environment are close to zero.
Therefore, thermodynamic efficiency of such an adiabatically operating motor can, in 
principle, be close to
the theoretical maximum of one. 
One can imagine, 
given already three particular
examples presented above, that it can be achieved, in principle, 
at the maximum of power, for arbitrary strong
dissipation. Design of the motor thus becomes crucially important. Such an ideal motor
can also be completely reversible. However, to arrive at the maximum thermodynamic
efficiency at a finite speed is a highly nontrivial matter indeed.

\subsubsection{Digression on a possibility of (almost) heatless classical computation}

We like to make now the following important digression.
In applications of these ideas to the physical principles of computation 
the above physical considerations
 mean the following: \\

Bitwise operation (bit ``0'' corresponds to one location of the potential
minimum and
bit ``1'' to another one -- let us assume that their energies are 
equal) does not required in principle any energy to finally 
dissipate (it can stored and reused during adiabatically slow change of potential). 
Physical computation can in principle be heatless, and it can be also completely 
reversible, at arbitrary dissipation. 
This is the reason why the original version of Landauer minimum principle allegedly 
imposed on computation (i.e. there is a minimum of $k_BT\ln 2$ of energy dissipated
 per  one bit of computation,
$0\to 1$, or $1\to 0$
required) was 
completely wrong, as recognized by late Landauer himself \cite{Landauer}, after 
Bennett \cite{Bennett}, Fredkin and Toffoli \cite{Fredkin} discovered how reversible
computation can be  done in principle \cite{Feynman}. Another currently 
popular version of Landauer principle 
in formulations that either one needs to spend  a minimum of $k_BT\ln 2$ of energy to destroy or 
erase one bit 
of information, or a minimum of $k_BT\ln 2$ heat is released by ``burning'' one bit of information
are also completely wrong. These two formulations plainly contradict, quite
generally, not analyzing any particular setup, to the second law of
thermodynamics, which in the differential form states that $d S\geq \delta Q/T $, i. e.
that the increase of entropy, or loss of information, $dI\equiv -dS/k_B\ln 2$ (a very fundamental
equality, or rather tautology of the physical information theory), is equal or 
exceeds the heat exchange with
the environment in the units of $T$. For an adiabatically isolated system,
 $\delta Q=0$, hence $dI \leq 0$, i.e. entropy can increase and information can diminish
 spontaneously, without any heat produced in the surrounding.
 This is just the second law of thermodynamics rephrased.
 As a matter of fact, $\delta Q=|dI|k_BT\ln2 $ is the maximal
 (not minimal!) amount of heat which can be produced by ``burning'' $dI$ bits of information.
 To create and store one bit of information one indeed needs to spend at least 
 $k_BT\ln 2$ amount of free energy
 at $T=const$, but not to destroy or erase it, in principle. 
 Information can be destroyed spontaneously, which can take, however, an infinite time. 
 Landauer principle  belongs scientifically to common fallacies. However, it presents a current hype
 in the literature at the same time. An ``economical'' reason for this is that current
 clock rate of  computer  processors
 stopped to increase below 10 GHz for over one decade 
 because of immense heat production. Plainly said, it is not possible to cool the processors anymore
 down if to further increase their rate, and the energy consumption becomes unreasonable. 
 We eagerly search for how to solve this
 severe problem.
 This problem is, however, a problem of the current design of these processors and our present 
 technology, which indeed provides severe thermodynamical limitations \cite{Kish02}. 
 However, it has anything
 in common with the Landauer principle as the heat is produced currently 
 by many orders of magnitude
 above the minimum of Landauer principle, which anyway should not be taken seriously as a rigorous  
 theoretical bound universally valid. Nevertheless, to operate at a finite speed is inevitably related
 to some heat losses. How to minimize them at a maximal speed? 
 This question is, however, clearly beyond the equilibrium thermodynamics. It belongs more to a kinetic
 theory. The minimum energy requirements are inevitably related to the question of how fast to compute.
 This presents currently an open unsolved problem.
 
 \subsection{Minimalist model of adiabatic pump}

\begin{figure}
\includegraphics[width=12cm,keepaspectratio]{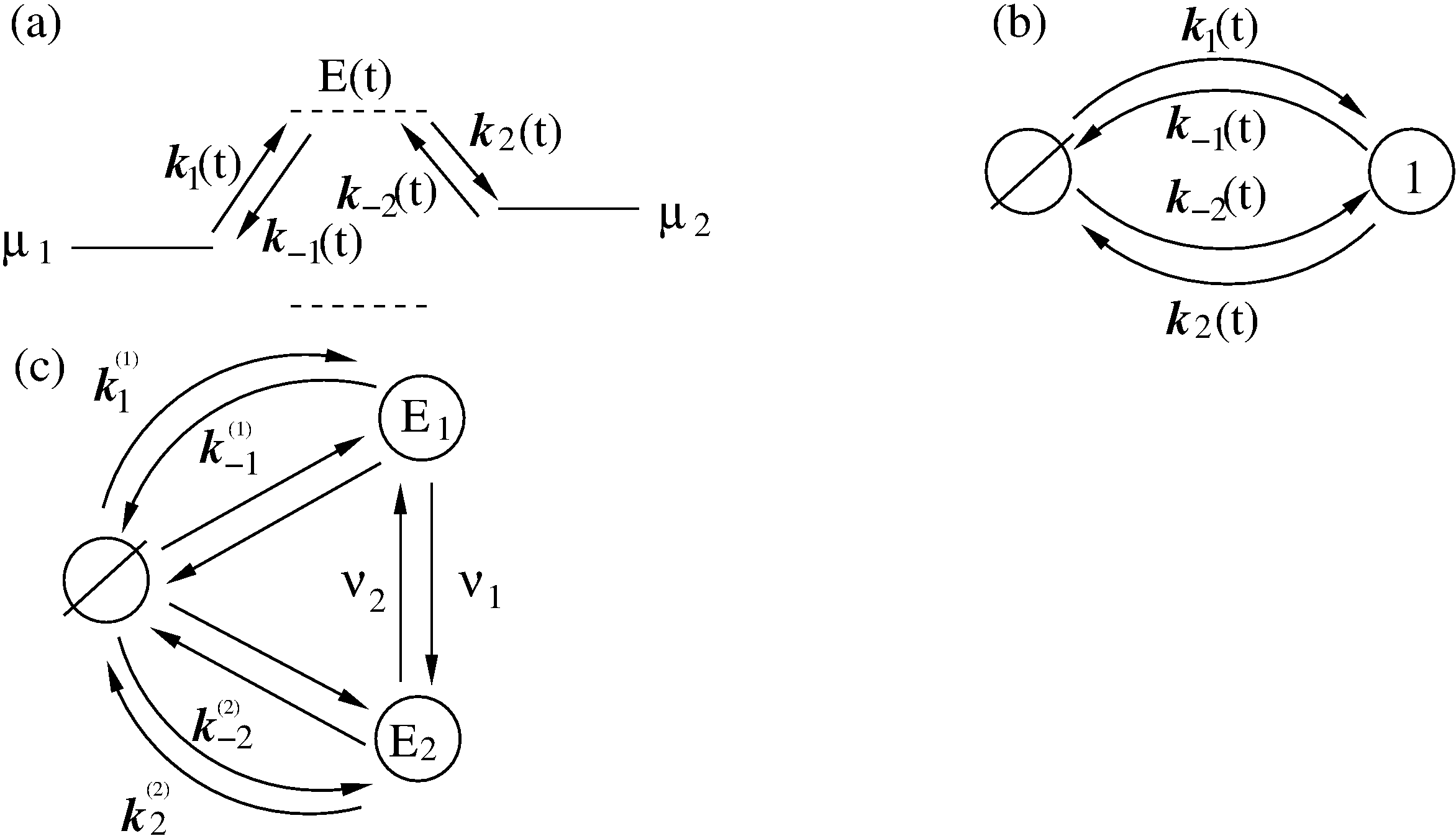} 
\caption{(a) Minimalist model of pump with one time dependent 
energy level $E(t)$ which can be used to  pump particles against a gradient of chemical potential $\Delta \mu$.
(b) Corresponding kinetic scheme with time-dependent rates and two states of the pump:
empty and filled with one transferred particle. Filling and emptying can occur from
either particle reservoirs, $\mu_1$ or $\mu_2$. (c) Equivalent kinetic scheme corresponding
to $E(t)$ having just two realizations with the transition rates $\nu_1$ and $\nu_2$.}
\label{Fig5}
\end{figure}
 
 Turning back to adiabatic operation of molecular motors or pumps, we shall analyze now 
 a minimalist  model based on time-modulation of the energy levels. 
 The physical background of the idea of adiabatic operation is sound. However, can
 it be realized in popular models featured by discrete energy levels?
 The minimalist model
 contains just one time-dependent energy level $E(t)$ and two constant energy levels
 corresponding to chemical potentials $\mu_1$ and $\mu_2$ of two baths of particles 
 between which the transport occurs. They must be considered as electrochemical
 potentials for charged particles, e.g. Fermi levels of electrons in two leads, or
 electrochemical potentials of transferred ions in two bath solutions separated
 by a membrane. 
 Pumping takes place when a time-modulation of
 $E(t)$ can be used to pump against $\Delta \mu=\mu_2-\mu_1>0$, see in Fig. \ref{Fig5}, (a).
 Here, both the energy level $E(t)$ and the corresponding rates $k_1(t)$ and $k_{-1}(t)$,
  $k_2(t)$ and $k_{-2}(t)$ are time-dependent. Their proper name would be rate constants,
  if they were time-independent.   Given sufficiently slow modulation and 
  fast equilibration
  at any instant $t$, one can assume \textit{local} equilibrium condition
  \begin{eqnarray}\label{local_eq}
  \frac{k_1(t)}{k_{-1}(t)}& = &\exp\left \{\beta  [E(t)-\mu_1] \right \},\nonumber \\
  \frac{k_2(t)}{k_{-2}(t)}& = &\exp\left \{-\beta  [E(t)-\mu_2] \right \}\,.
\end{eqnarray}
Notice, that this condition is not universally valid. It can be violated by fast fluctuating
fields, see in \cite{GoychukAP05} and references cited therein, 
for a plenty of examples, and an approach beyond this restriction within
a quantum-mechanical setting. Rates are generally retarded functionals of energy levels fluctuations
and not functions of instant energy levels. However, local equilibrium can be 
a very good approximation. 
Fig. \ref{Fig5}, (b) rephrases the transport
process in  Fig. \ref{Fig5}, (a) in terms of the states of the pump: empty, state $\emptyset$, and filled
with one transferred particle, state $1$. The former state is populated with probability $p_0(t)$,
and the latter one with probability $p_1(t)$, $p_0(t)+p_1(t)=1$. The empty level can be filled
with rate $k_1(t)$ from the left bath level $\mu_1$, and with rate $k_{-2}(t)$ from the 
right bath level $\mu_2$. The filling flux is thus $j_f=(k_1+k_{-2})p_0$. Moreover, it can be 
emptied with rate $k_2(t)$ to $\mu_2$, and with rate $k_{-1}(t)$ to $\mu_1$.
The corresponding master equations reduce to a single relaxation equation
because of probability conservation:
\begin{eqnarray}\label{relax}
\dot p_0(t)& =&-[k_1(t)+k_{-2}(t)]p_0(t)+[k_2(t)+k_{-1}(t)]p_1(t)\nonumber \\
& = & -\Gamma(t)p_0(t)+R(t)\;,
\end{eqnarray}
where 
\begin{eqnarray}
\Gamma(t)=k_1(t)+k_{-1}(t)+k_{-2}(t)+k_{2}(t)\;,
\end{eqnarray}
and 
\begin{eqnarray}
R(t)=k_{-1}(t)+k_{2}(t)\;.
\end{eqnarray}
The instant flux between the levels $\mu_1$ and $E(t)$ is
\begin{eqnarray}
I_1(t)=k_1(t)p_0(t)-k_{-1}(t)p_1(t)\;,
\end{eqnarray}
and 
\begin{eqnarray}
I_2(t)=k_2(t)p_1(t)-k_{-2}(t)p_0(t)\;,
\end{eqnarray}
between the levels $E(t)$ and $\mu_2$. Clearly,  the time averages 
$\bar I_1=\lim_{t\to\infty}(1/t)\int_0^{t}I_1(t')dt'$ and 
$\bar I_2=\lim_{t\to\infty}(1/t)\int_0^{t}I_1(t')dt'$ must coincide, $\bar I_1=\bar I_2=I$,
because particles cannot accumulate on the level $E(t)$.

First, we show that pumping is impossible within the approximation of quasi-static rate, i.e.
when the rates are considered to be constant at a frozen time instant and one solves the problem
within this approximation. Indeed, in this case for a steady-state flux which is instant function
of time we obtain:
\begin{eqnarray}\label{quasi-static}
I(t)& = &\frac{k_1(t)k_2(t)-k_{-1}(t)k_{-2}(t)}{\Gamma(t)}\nonumber \\
& = & \frac{k_1(t)k_2(t)}{\Gamma(t)}\left [ 1-e^{\Delta \mu/(k_BT)}\right ]
\;,
\end{eqnarray}
where in the second line we used (\ref{local_eq}). Clearly
for $\Delta \mu>0$, $I(t)<0$ at any $t$. Averaging over time yields, 
\begin{eqnarray}\label{quasi-static-av}
\bar I_{qs}=\bar I_f\left [ 1-e^{\Delta \mu/(k_BT)}\right ]\;,
\end{eqnarray}
with $\bar I_f=\lim_{{\cal T}\to\infty}(1/{\cal T})\int_0^{\cal T} k_1(t)k_2(t)dt/\Gamma(t)$.
The current flows always from higher $\mu_2$ to
lower $\mu_1$. The same will happen for any number of intermediate levels $E_i(t)$ within such
an approximation. 

\subsection{Origin of pumping}

One can, however, easily solve Eq. (\ref{relax}) for arbitrary $\Gamma(t)$ and $R(t)$:
\begin{eqnarray}\label{exact}
p_0(t)=p_0(0)e^{-\int_0^t\Gamma(t')dt'}+\int_0^t dt'R(t')e^{-\int_{t'}^t\Gamma(t'')dt''}\;.
\end{eqnarray}
The first term vanishes in the limit $t\to\infty$ and
a formal expression for steady state averaged flux $\bar I$ can be readily written,
\begin{eqnarray}
\bar I=\lim_{\cal T\to\infty}\frac{1}{\cal T}
\int_0^{\cal T}dt[k_1(t)+k_{-1}(t)]\int_0^t dt'R(t')e^{-\int_{t'}^t\Gamma(t'')dt''}-\bar k_{-1}\;,
\end{eqnarray}
where $\bar k_{-1}$ is time-averaged $k_{-1}(t)$.
However,
to evaluate it for some particular protocols of energy and rates modulation is generally a rather 
cumbersome task. The fact that pumping is possible is easy to understand making the 
following protocol of energy level and rates modulation: (step 1) energy level $E(t)$
goes down, $E(t)<\mu_1$, with an increasing prefactor in $k_{\pm 1}(t)$ (left gate opens),
and sharply decreasing prefactor in $k_{\pm 2}(t)$ (right gate is closed), a particle
enters pump from the left; (step 2)
energy level $E(t)$ goes up, $E(t)>\mu_2$, and prefactor in $k_{\pm 1}(t)$ sharply drops,
the left gate closes and the right one remains closed; (step 3) the right gate opens and
the particle leaves to the right; (step 4) the right gate closes, the energy level $E(t)$ goes
down and the left gate opens, so that the initial position in three-dimensional parameter space 
(two prefactors and one energy level) is repeated, and one cycle is completed.
The general idea of ionic pump with two intermittently opening/closing gates has in fact been 
suggested long time ago \cite{Jardetzky}.

Some general results can be obtained within this model for adiabatic slow modulation and related to
an adiabatic geometric Berry phase $b(t)$, the origin of which can be understood, per analogy 
with a similar approach used
to solve Schroedinger equation in quantum mechanics for adiabatically modulated 
quasi-stationary energy levels \cite{Anandan},
by making the following Ansatz to solve Eq. (\ref{relax}): 
$p_0(t)=e^{i b(t)}R(t)/\Gamma(t) + c. c.$. Making a loop  in a two-dimensional space
of parameters adds or subtracts 
$2\pi$ to $b(t)$. Furthermore, an additional related contribution, pumping current, 
appears in addition to one in (\ref{quasi-static-av}), with averaging done over one cycle period. 
This additional contribution is proportional to the cycling rate $\omega$,  
see in \cite{Sinitsyn} for detail. However, it is small and cannot override
one in (\ref{quasi-static-av}) consistently with the adiabatic modulation assumptions. Hence,
adiabatic pumping against any substantial bias $\Delta \mu>0$ is not possible within this model. 
This indeed can easily be understood by making a sort of adiabatic approximation in
Eq. (\ref{exact}), $\int_{t'}^{t}\Gamma(t'')dt''\approx \Gamma(t)(t-t')$, and doing an
integration by parts therein, so that in the long time limit 
$p_0(t)\approx R(t)/\Gamma(t)+\delta p_0(t)$,
where $\delta p_0(t)\propto \dot R(t)\propto \omega$. The first term leads to (\ref{quasi-static-av}),
and the second term corresponds to a small perturbative pump current, which vanishes as $\omega\to 0$. 
This  pump current can be observed only for $\Delta \mu=0$, where $\bar I_{qs}=0$. 
Hence, thermodynamic efficiency of this 
pump is close to zero in adiabatic pumping regime. 

Moreover, for realistic molecular pumps driven e.g. by energy of ATP hydrolysis, the adiabatic
modulation is difficult if possible to realize. A sudden modulation of the energy levels,
i.e. a power stroke,
 when the energy
levels jump to new discrete positions, is more relevant, especially on a single-molecular level.

\subsection{Efficient non-adiabatic pumping }

The cases,
where $E(t)$ takes on discrete values being a continuous time semi-Markovian process 
can be handled
differently. Especially simple is a particular case with $E(t)$ taking just two values 
$E_1$ and $E_2>E_1$ with transition rates $\nu_1$, and $\nu_2$ between those. Then, the transport 
scheme in Fig. \ref{Fig5}, (b) can be rephrased as one in Fig. \ref{Fig5}, (c) with rate constants
$k_{j}^{(i)}$ for the transitions to and from the energy levels $E_i$, $i=1,2$, $j=1,2,-1,-2$, 
and 
\begin{eqnarray}\label{local_eq2}
  \frac{k_{1}^{(i)}}{k_{-1}^{(i)}}& = &\exp\left \{\beta  [E_i-\mu_1] \right \},\nonumber \\
  \frac{k_{2}^{(i)}}{k_{-2}^{(i)}}& = & \exp\left \{-\beta  [E_i-\mu_2] \right \} \,.
\end{eqnarray}
Now we have three populations, $p_0$ of empty state, $p_1$ of level $E_1$, and $p_2$ of level $E_2$.
The steady state flux can be calculated as $I=\nu_1 p_1^{(st)}-\nu_2 p_2^{(st)}$, where $p_{1,2}^{(st)}$
are steady state populations. Straightforward, but somewhat lengthy calculations yield
\begin{eqnarray}\label{flux}
I=\frac{\nu_1\left (k_{1}^{(1)}+k_{-2}^{(1)}\right)\left (k_{-1}^{(2)}+k_{2}^{(2)}\right)-
\nu_2\left (k_{1}^{(2)}+k_{-2}^{(2)}\right)\left (k_{-1}^{(1)}+k_{2}^{(1)}\right)}
{\sum_i k_{i}^{(1)}\sum_j k_{j}^{(2)}+\nu_1\left (\sum_j k_{j}^{(2)}+k_{1}^{(1)}+k_{-2}^{(1)}\right)
+\nu_2\left (\sum_j k_{j}^{(1)}+k_{1}^{(2)}+k_{-2}^{(2)}\right)}  \;.
\end{eqnarray}
From the structure of this equation it is immediately clear that the flux can be positive for positive
$\Delta \mu$ (real pumping), e.g. by considering the limit:
$k_{1}^{(1)}\gg k^{(1)}_{-2}$, $k_{-1}^{(2)}\ll k_{2}^{(2)}$, $k_{1}^{(2)}\ll k_{-2}^{(2)}$,
$k_{2}^{(1)}\ll k_{-1}^{(1)}$, and 
$\nu_1\gg \nu_2$. Physically, it is obvious when
$E_1<\mu_1$, and $E_2>\mu_2$, together with $k_{1}^{(1)}\gg k_{-2}^{(1)}$ (i.e. the level $E_1$
is easily filled from $\mu_1$, but not from $\mu_2$ because e.g. of  a large barrier on
the right side -- the entrance of pump is practically closed from the right), 
and $k_{2}^{(2)}\gg k_{-1}^{(2)}$
(i.e. the particle easily goes from $E_2$ to $\mu_2$ and cannot go back to $\mu_1$ because
the left entrance is now almost closed). Under these conditions, also $k_{1}^{(2)}\ll k_{-2}^{(2)}$
and $k_{2}^{(1)}\ll k_{-1}^{(1)}$ are well justified. Hence, we obtain for the pumping rate
\begin{eqnarray}\label{flux_approx}
I\approx \frac{\nu_1 k_{1}^{(1)}k_{2}^{(2)}}
{k_{1}^{(1)}k_{2}^{(2)}+\nu_1  \left (k_{1}^{(1)}+k_{2}^{(2)} \right )}  \;.
\end{eqnarray}
This expression looks like a standard Michaelis-Menthen rate of enzyme operation,
which is customly used in biophysics \cite{Nelson} for modeling molecular motors
and pumps.
Elevation of $E(t)$ level from $E_1$ to $E_2$ can be effected e.g. by ATP binding in case ionic
pumps, with $\nu_1\propto c_{ATP}$, where $c_{ATP}$ is ATP concentration. This is a simplest basic
model for pumps. From (\ref{flux_approx}) it follows that $I\approx \nu_1$ at $\nu_1 \tau \ll 1$, where
$\tau=1/k_{1}^{(1)}+1/k_{2}^{(2)}$ is the sum of filling and emptying times, and it reaches
the maximal pumping rate $I_{\rm max}\approx 1/\tau$, for $\nu_1 \tau \gg 1$. Thermodynamic efficiency
of such a pump is $R=\Delta \mu/\Delta E$, where $\Delta E=E_2-E_1$ is energy invested in 
pumping. Derivation of approximate Eq. (\ref{flux_approx}) requires
that $\exp(\epsilon_{1,2}/k_BT)\gg 1$, where $\epsilon_1=\mu_1-E_1$, and $\epsilon_1=E_2-\mu_2$,
which is well satisfied already for $\epsilon_{1,2}>2k_BT$. Hence 
$R=\Delta \mu/(\Delta \mu +\epsilon_1+\epsilon_2)$ can be close to one for a large
$\Delta \mu \gg \epsilon_1+\epsilon_2$. Take for example $\Delta E =20\;k_BT_r\approx 0.5$ eV,
which corresponds to a typical energy released by ATP hydrolysis. Then, for $\Delta \mu=0.4$ eV
and $\epsilon_1=\epsilon_2=2k_BT_r\approx 0.05$ eV, $R=0.8$. Notice that a typical thermodynamic
efficiency of Na-K pump is about $R\approx 0.75$.
Such a non-adiabatic pumping can thermodynamically
 be highly efficient indeed
with small heat losses.
One should remark, however, that the question on whether or not
the efficiency at the maximum of power, $P_W=I\Delta \mu$, can be larger than 
one-half or even approach  one
within this generic model is not that simple.
To answer this question, one cannot neglect backward transport, especially 
when $\Delta \mu$ becomes close to $\Delta E$ ($P_W(\Delta \mu=\Delta E)=0$), 
and one has to specify a concrete model for the rates 
in the exact result (\ref{flux}). In the case of an electronic pump, like one used by nature
in nitrogenase enzymes this can be quantum tunneling rates, see in \cite{GoychukMolSim06}, like
Marcus-Levich-Dogonadze rate above. Moreover, imposing intermittently in time a very high barrier either
on the left, or on the right can physically correspond to interruption 
of electron tunneling pathway due to
ATP-induced conformational changes, i.e. to
modulation of tunnel coupling $V_{\rm tun}(t)$ synchronized with modulation
of $E(t)$, as it does occur in nitrogenase. This question of efficiency at maximum power will be analyzed elsewhere else
in detail, both for classical and quantum rate models.

To summarize this section, the idea of adiabatic operation of molecular machines 
is sound. It should be pursued further.
However, the known simplest adiabatic pump operates in fact at nearly zero thermodynamical efficiency, 
while a power stroke mechanism can be highly efficient within the same model. It seems obvious that
in order to realize a thermodynamically  efficient adiabatic pumping, a gentle operation of
molecular machine without erratic jumps,
 one needs a continuum of states,
or possibly many states depending continuously on an external modulation parameter.
A further research is thus highly desirable and needed.

\section{How can biological molecular motors operate highly efficiently
in highly dissipative viscoelastic environments?}

As it has been clarified above, Brownian motors can work highly efficiently
in dissipative environments causing arbitrary strong viscous friction acting on motor. 
This corresponds to the case of normal diffusion, $\langle [\delta x(t)]^2\rangle =2D t$, 
in a force-free case. In a crowded environment of biological cells, diffusion can be, however,
anomalously slow, $\langle [\delta x(t)]^2\rangle =2D_\alpha t^\alpha/\Gamma(1+\alpha)$, where
$0<\alpha<1$ is power law exponent of subdiffusion, and $D_\alpha$ is 
subdiffusion coefficient \cite{Barkai,Hofling}.
There is a huge body of growing experimental evidence for subdiffusion of particles of various sizes,
from $2-3$ nm (typical for globular proteins) \cite{Guigas,Saxton} to $100-500$ nm 
\cite{Seisenberg,Golding,Tolic,Jeon11,Parry}
(typical for various endosomes), both
in living cells, and in crowded polymer and colloidal solutions (complex fluids) 
physically resembling cytoplasm. There are many theories developed to explain such a 
behavior \cite{Barkai,Hofling}.
One is based on natural viscoelasticity of such complex liquids, see \cite{GoychukACP12,Waigh} 
for a review  and detail. It has a deep dynamical foundation (see above). 
Viscoelasticity which leads to above subdiffusion corresponds to
a power law memory kernel $\eta(t)=\eta_\alpha t^{-\alpha}/\Gamma(1-\alpha)$ in 
Eqs. (\ref{GLE}), (\ref{FDR}), where $\eta_\alpha$ is fractional friction coefficient,
which is related to $D_\alpha$ by the generalized Einstein relation, $D_\alpha=k_BT/\eta_\alpha$.
Using the notion of fractional Caputo derivative, the dissipative term in Eq. (\ref{GLE})
can be abbreviated as $\eta_\alpha d^\alpha x/dt^\alpha$, where the fractional derivative
operator $d^\alpha f(t)/dt^\alpha$ acting on arbitrary function $f(t)$ is just 
defined by this abbreviation. The corresponding GLE is named fractional Langevin equation (FLE).
Its solution yields the above subdiffusion 
scaling exactly in the inertialess limit,  
$M\to 0$, corresponding precisely to the fractional Brownian motion \cite{GoychukPRL07a,GoychukACP12}, 
or asymptotically otherwise. The transport in the case of a constant force $f_0$ applied
is also subdiffusive, $\langle \delta x(t)\rangle =(f_0/\eta_\alpha) t^\alpha/\Gamma(1+\alpha)=
(f_0/(2k_BT))\langle [\delta x(t)]^2\rangle$.
These results correspond exactly to sub-Ohmic  model of the spectral density of thermal bath
\cite{WeissBook}, 
$J(\omega)=\eta_\alpha \omega^\alpha$, within the dynamical approach to generalized Brownian motion.
They can be easily understood if to do ad hoc Markovian approximation to
the memory kernel, which yields a time-dependent viscous friction, $\eta_M(t)\dot x(t)$
with $\eta_M(t)\propto t^{1-\alpha}$. It diverges, $\eta_M(t)\to\infty$, when $t\to\infty$, which
leads to subdiffusion and subtransport within this Markovian approximation.
Such an approximation can, however, be very misleading in other aspects \cite{GoychukPRE15}.
Nevertheless, it provokes the question: How can molecular motors, such as kinesin,
 work very efficiently in such media characterized by virtually infinite friction, interpolating
 in fact between simple liquids and solids? Important to mention, in any fluid-like
 environment the effective macroscopic friction, $\eta_{\rm eff}=\int_0^{\infty}\eta(t)dt$,
 must be finite. Hence, a memory cutoff time $\tau_{\rm max}$ must exist, so that 
 $\eta_{\rm eff}\propto \eta_{\alpha}\tau_{\rm max}^{1-\alpha}$. In real life, 
 $\tau_{\rm max}$ can be as large as minutes, or even longer than hours. Hence, on a shorter time 
 scale and on a corresponding spatial mesoscale it is subdiffusion, characterized by
 $\eta_{\alpha}$, which can physically be
 relevant indeed and not the macroscopic limit of normal diffusion characterized by
 $\eta_{\rm eff}$. This observation opens a way for multi-dimensional 
 Markovian embedding of subdiffusive processes
 with long range memory upon introduction of a \textit{finite} number $N$ of
  auxiliary stochastic variables. It is based on a Prony series expansion of power-law memory kernel
 into a sum of exponentials, $\eta(t)=\sum_{i=1}^N k_i \exp(-\nu_i t)$, with
 $\nu_i=\nu_0/b^{i-1}$ and $k_i\propto \nu_i^\alpha$, which can be made very accurate numerically
 (this is controlled by the scaling parameter $b$), and apart from 
 $\tau_{\rm max}=\tau_{\rm min}b^{N-1}$ possesses also a short cutoff $\tau_{\rm min}=1/\nu_0$.
 The latter one naturally emerges in any condensed medium beyond continuous medium approximation
 because of real atomistic nature. In numerics, it can be made of the order
 of time integration step. Hence, it does not matter even within the continuous medium approximation.
 Even with a moderate $N\sim 10-100$ (number of auxiliary degrees of freedom) Markovian 
 embedding can be done
 for any realistic time scale of anomalous diffusion with sufficient accuracy \cite{GoychukPRE09,
 GoychukACP12}.
 A very efficient numerical approach based on the corresponding Markovian embedding has 
 been developed for subdiffusion in Refs. \cite{GoychukPRE09,GoychukACP12}, and for 
 superdiffusion ($\alpha>1$)
  in Refs. \cite{SieglePRE10,SieglePRL10,SiegleEPL11}. The idea of Markovian 
  embedding is also very natural in view of that
 any non-Markovian GLE dynamics presents a low-dimensional projection of a hyper-dimensional
 singular Markovian process described by dynamical equations of motion with random initial
 conditions. This fact is immediately clear from a well-known dynamical derivation of GLE
 reproduced above. Somewhat surprising is, however, that
 not so many $N\sim 10-20$ are normally sufficient in practical applications.

The action of a motor on subdiffusing cargo can be simplistically modeled (a simplest possible theory)
by a random force $f(t)$ alternating its direction, 
when the motor steps on a random network of
cytoskeleton \cite{Caspi}. The driven cargo follows a diffusional 
process $\langle [\delta x(t)]^2\rangle_{t'} \propto
t^\beta$, with some exponent $\beta$, if to make a trajectory averaging of squared displacements
$\delta x(t|t')=x(t+t')-x(t')$  over sliding $t'$. Within such a modeling
$\beta$ clearly cannot exceed $2\alpha$ \cite{BrunoPRE}, 
that corresponds to subtransport which alternates its
direction in time. Hence, for $\alpha<0.5$ no cargo superdiffusion ($\beta>1$) could not be caused by
motors within such a simple approach. However, experiments show \cite{Robert,Harrison} that freely subdiffusing cargos 
(e.g. $\alpha=0.4$ \cite{Robert,Bruno11}) can superdiffuse , 
when they are driven by motors also
for $\alpha<0.5$ (e.g. $\beta=1.3$ for $\alpha=0.4$ \cite{Robert}) ). Hence, a more appropriate 
modeling of the transport by molecular motors in viscoelastic
environments is required. It has been developed quite recently in Refs. \cite{PLoSONE14,PCCP14,PhysBio15}, 
by generalizing
pioneering works on subdiffusive rocking 
\cite{ChemPhys10,GoychukACP12,PRE12a,PRE13,MMNP13}, and 
flashing \cite{NJP12} ratchets.

\begin{figure}
\includegraphics[width=12cm,keepaspectratio]{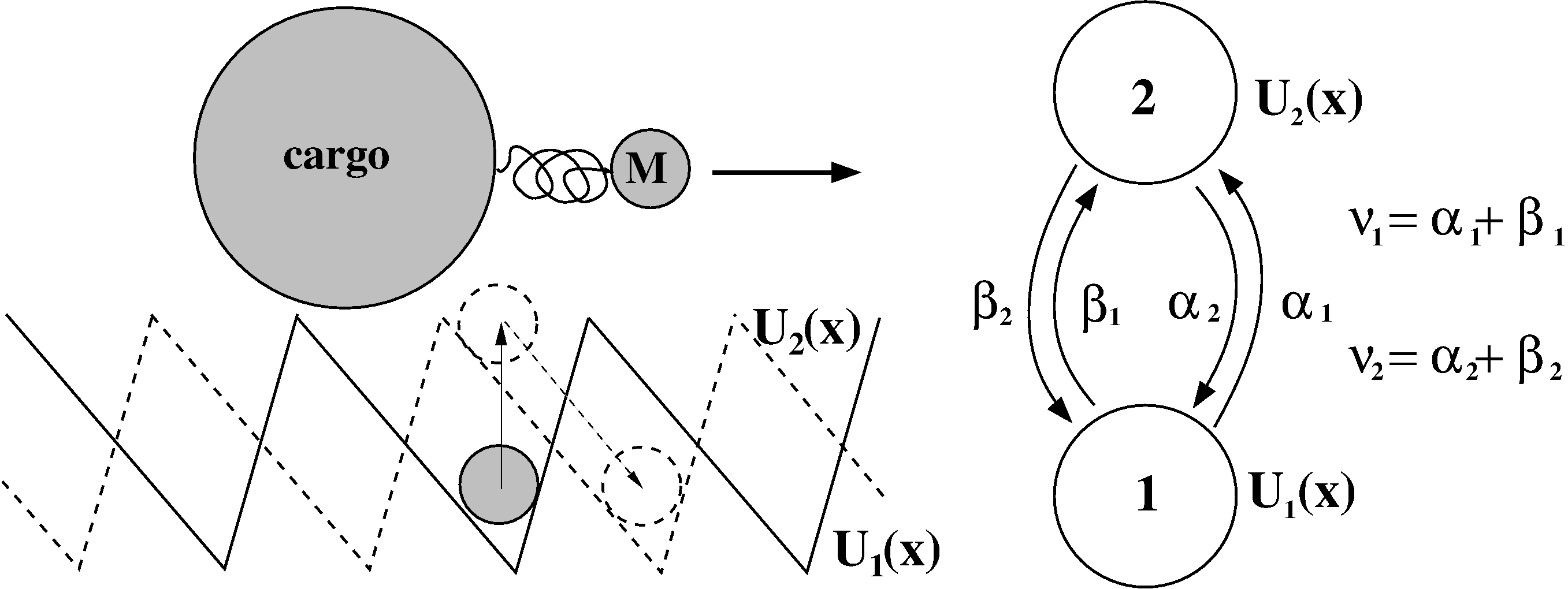}
\caption{Motor pulling cargo on an elastic linker. The motor can be trapped in a flashing
periodic potential (here, two realizations shifted by a half of spatial period are shown).
These fluctuations are caused and driven by conformational fluctuations of the motor 
protein. The minimalist pertinent two-state cyclic model of the corresponding biochemical
enzymatic cycle is shown on the right. Mechanical motion, induced by cycling,
exerts a back influence on cycling via spatially dependent transition rates.
This can cause anomalously slow enzyme kinetics, which cannot be characterized
by a turnover rate, in viscoelastic environments \cite{PhysBio15}. 
}
\label{Fig6} 
\end{figure}

Viscoelastic effects should by considered on the top of viscous Stokes friction caused by
the water component of cytosol. Then, a basic 1d model for a large cargo (20-500 nm) pulled
by a much smaller motor (2-10 nm) on an elastic linker, cf. Fig. \ref{Fig6},
 can be formulated as follows \cite{PhysBio15}
\begin{eqnarray}
\label{model1a}
\eta_c \dot y &= &- \int_{0}^t\eta(t-t')\dot{y}(t')dt'-\frac{k_L(y-x)}{1-(y-x)^2/r_{\rm max}^2}
+\xi_c(t)+\xi(t),\\
 \eta_m\dot{x} &=&
\frac{k_L(y-x)}{1-(y-x)^2/r_{\rm max}^2}-\frac{\partial}{\partial x}U(x,\zeta(t))-f_L +\xi_m(t).
\label{model1b}
\end{eqnarray}
This presents a generalization of a well-known model of molecular motors \cite{Julicher,AstumianBier,
Parmeggiani} by coupling motor to a subdiffusing cargo on an elastic linker.
Here, both the motor (coordinate $x$) and the cargo (coordinate $y$) are subjected to independent thermal white
noises of the environment, $\xi_m(t)$, and $\xi_c(t)$, respectively, which obey the corresponding
FDRs. The both particles are overdamped and characterized by the Stokes frictional forces with frictional
constants $\eta_m$, and $\eta_c $. In addition, on cargo acts viscoelastic frictional force
characterized by the memory kernel discussed above (fractional friction model) and the corresponding
stochastic thermal force $\xi(t)$ with algebraically decaying correlations. It obeys
a corresponding FDR. The motor can pull cargo on an elastic linker with spring constant $k_L$
(small extensions) and maximal extension length $r_{\rm max}$ (the so-called finite extension
nonlinear elastic (FENE) model \cite{FENE} is used here). The motor (kinesin) is bound to a microtubule and can move
along it in a periodic potential $U(x+L,\zeta(t))=U(x,\zeta(t))$ reflecting microtubule spatial period
$L$, and do a useful work against a loading force $f_L$ directed against its motion
caused by cyclic conformational fluctuations $\zeta(t)$. Microtubule is a polar periodic 
structure with a periodic but asymmetric distribution of positive
and negative charges (overall charge is negative) \cite{Baker}. Kinesin is also charged and its charge fluctuates
upon binding negatively charged ATP molecules and dissociation of the products of ATP hydrolysis.
This leads to dependence of the binding potential on the conformational variable $\zeta(t)$.
Given two identical heads of kinesin, the minimalist model is to assume that there are only two
conformational states of the motor (this is a gross oversimplification, of course) with 
$U_{1,2}(x):=U(x,\zeta_{1,2})$, and $U_{1}(x+L/2)=U_{2}(x)$ as an additional symmetry condition,
so that a half-step $L/2$ is associated with conformational fluctuations $1\to 2$, or $2\to 1$.
During one cycle $1\to  2\to 1$ in the forward direction with the rates $\alpha_1(x)$ and
$\beta_2(x)$, one ATP molecule is hydrolyzed, whereas if this cycle is reverted in the backward 
direction with the rates  $\beta_1(x)$ and $\alpha_2(x)$, see Fig. \ref{Fig6}, one ATP molecule
is synthesized. 
The dependence of chemical transition rates on the position $x$ through the 
potential $U_{1,2}(x)$ reflects a two way mechano-chemical coupling.
It it is capable to incorporate allosteric effects
which indeed can be very important for optimal operation of molecular machines \cite{Cheng}. 
Such effects can possibly emerge, for example, because the probability of binding 
of ATP molecule (substrate) to
kinesin motor or release of products can be influenced by electrostatic potential of the microtubule.
In the language 
of Ref. \cite{Cheng} this corresponds to an information ratchet mechanism to distinguish it
from the energy ratchet where the rates of potential switches do not depend on the motor states
(no feedback) and are fixed. Such an allostery can be used to create highly efficient molecular
machines \cite{Cheng}.
In accordance with general principles of nonequilibrium thermodynamics applied
to cyclic kinetics \cite{Hill}
\begin{eqnarray}
\frac{\alpha_1(x)\beta_2(x)}{\alpha_2(x)\beta_1(x)}=\exp[|\Delta \mu_{\rm ATP}|/(k_BT)],
\end{eqnarray}
for any $x$, where $|\Delta \mu_{\rm ATP}|$ is the free energy released in ATP hydrolysis
and used to drive one complete cycle in forward direction. It 
can be satisfied, e.g.,  by choosing
\begin{eqnarray}
\frac{\alpha_1(x)}{\alpha_2(x)}& = & \exp[(U_1(x)-U_2(x)+|\Delta \mu_{\rm ATP}|/2)/(k_BT)], \nonumber \\
\frac{\beta_1(x)}{\beta_2(x)}& = & \exp[(U_1(x)-U_2(x)-|\Delta \mu_{\rm ATP}|/2)/(k_BT)] .
\end{eqnarray}
The total rates 
\begin{eqnarray}
\nu_1(x)=\alpha_1(x)+\beta_1(x), \nonumber \\
\nu_2(x)=\alpha_2(x)+\beta_2(x)
\end{eqnarray}
of the transitions between two energy profiles must satisfy
\begin{eqnarray}
\frac{\nu_1(x)}{\nu_2(x)}=\exp[(U_1(x)-U_2(x))/(k_BT)]
\end{eqnarray}
at thermal equilibrium. This is condition of the thermal detailed balance, 
where the dissipative fluxes vanish
both in the transport direction and within the conformational space of motor,
at the same time \cite{Julicher,AstumianBier}. 
It is obviously satisfied 
for $|\Delta \mu_{\rm ATP}|\to 0$. Furthermore, on symmetry grounds, not
only $\alpha_{1,2}(x+L)=\alpha_{1,2}(x)$, 
$\beta_{1,2}(x+L)=\beta_{1,2}(x)$, but also, 
$\alpha_{1}(x+L/2)=\beta_{2}(x)$ and $\alpha_{2}(x+L/2)=\beta_{1}(x)$.
It should be emphasized that such linear motors as kinesin I or II  work only one way:
they utilize chemical energy of ATP hydrolysis for doing mechanical work. They cannot 
operate in reverse on average, i.e. to use mechanical work in order to produce
 ATP in a long run, even if a two way mechano-chemical coupling
 can provide  such an opportunity in principle. This is very different from such rotary motors as 
 F0F1-ATPase which is completely reversible and can operate in two opposite directions.
 Allosteric effects can also play a role to provide such a directional
 asymmetry in the case of kinesin
 motors.  Allostery should be considered as generally important for a proper design 
 of various motors best suited for different tasks.
 
 For kinesins, neither cargo nor external force $f_L$ should explicitly influence
 the chemical rate dependencies on the mechanical coordinate $x$. 
 This 
leaves still some freedom in use of different models of rates. One possible 
choice is \cite{PhysBio15}
\begin{eqnarray}\label{nu1}
 \nu_1(x)& = &\alpha_1(x)+ \alpha_1(x+L/2)\exp[-(U_2(x)-U_1(x)
 + |\Delta \mu_{\rm ATP}|/2)/(k_BT)], \nonumber \\
 \nu_2(x)& = & \alpha_1(x)\exp[-(U_1(x)-U_2(x)+|\Delta \mu_{\rm ATP}|/2)/(k_BT)] 
 + \alpha_1(x+L/2)\;, \label{nu2}
\end{eqnarray}
with $\alpha_1(x)=\alpha_1$ within some
$\pm\delta/2$ neighborhood of the minimum of potential $U_1(x)$ and  zero otherwise.
Correspondingly, the rate $\beta_2(x)=\alpha_1$ within 
$\pm\delta/2$ neighborhood of the minimum of potential $U_2(x)$.
The rationale behind this choice is that these rates correspond to lump 
reactions of ATP binding and hydrolysis, and if to choose the amplitude of
the binding potential to be about $|\Delta \mu_{\rm ATP}|$, and a sufficiently large 
$\delta$, the rates $\nu_{1,2}(x)$
can be made almost independent of the position of motor 
along microtubule \cite{PhysBio15}, considering allosteric effects 
to be almost negligible in this particular respect. 
This allows to compare this model featured
by bidirectional mechano-chemical coupling with a corresponding flashing energy 
ratchet model,
where the switching rates between two potential realizations are spatially-independent 
constants, $\nu_1=\nu_2=\alpha_1$. 
The latter model has been developed in Ref. \cite{PCCP14}. 
Notice that
even for reversible F1-ATPase motors such an energy ratchet model can provide very reasonable
and experimentally relevant results \cite{Perez}.
Moreover, if the linker is very rigid, $k_L\to\infty$, one can exclude the
dynamics of cargo and to consider one compound particle with a renormalized Stokes friction
and the same algebraically decaying memory kernel  and moves \textit{subdiffusively} in
a flashing potential. Such an anomalous diffusion molecular motor model has been proposed
and investigated in Ref. \cite{PLoSONE14}. The main results of \cite{PLoSONE14}, which were
confirmed and further generalized in  \cite{PCCP14,PhysBio15}, create the following emerging
coherent picture of molecular motors pulling  subdiffusing (when free) cargos in viscoelastic 
environment of living cells. First, if normally diffusing (when free) motor is coupled to 
subdiffusing cargo it will be eventually enslaved by the cargo and also subdiffuse  \cite{PCCP14}. 
However,
when the motor is bounded to microtubule, it can be guided by the binding potential fluctuations, which are
eventually induced by its own cyclic conformation dynamics driven by the free energy released in ATP 
hydrolysis. It slides towards a new potential minimum after each potential change, or can 
fluctuationally escape to another minimum, see Fig. \ref{Fig6} to realize this. Large binding 
potential amplitude $U_0\gg k_BT$ (should exceed $10-12\,k_BT$, see Fig. 6
and a corresponding discussion  in \cite{PCCP14} to understand why) makes the motor strong. For a large
$U_0$ the probability to escape is small, and the motor will typically slide down to a new minimum
and its mechanical motion along microtubule will be completely synchronized with potential
flashes and conformational cycles. It steps then (stochastically but unidirectionally) 
to the right in Fig. \ref{Fig6} with mean velocity ${\rm v}=L\alpha_1/2$. In such a way, using
a power stroke like mechanism a strong
motor like kinesin II (with stalling force $F \sim 6-8$ pN) can completely
 beat subdiffusion 
and transport very efficiently even subdiffusing (when free) cargos. This requires, however, that flashing occurs
slower than relaxation. The larger the cargo the larger is also the fractional friction 
coefficient $\eta_\alpha$, and slower relaxation. The relaxation is algebraically slow. However,
it can be sufficiently fast in absolute terms on the time scale $1/\alpha_1$, so that mechanism is realized
for sufficiently small cargos. The results of \cite{PLoSONE14,PCCP14,PhysBio15} indicate that
smaller cargos, $20-100$ nm, will typically be transported by strong kinesin motors quite
normally, $\langle y(t)\rangle \sim \langle x(t)\rangle \propto t^{\alpha_{\rm eff}}$, with
$\alpha_{\rm eff}=1$, at typical motor turnover frequencies $\nu=\alpha_1/2\sim 1-200$ Hz,
provided that $f_L=0$. This already explains why the diffusional exponent $\beta\sim 2
\alpha_{\rm eff} $ can be larger than $2\alpha$. However, for larger cargos, $100-300$ nm, 
larger turnover frequencies
and when the motor works in addition against a constant loading force $f_L$, anomalous transport
regime emerges with $\alpha\leq \alpha_{\rm eff} \leq 1$. Clearly, when $f_L$ approaches
the stalling  force $F$ the transport becomes anomalous.
The effective transport exponent
$\alpha_{\rm eff}$ is thus essentially determined by binding potential strength, motor operating
frequency, cargo size, and loading force, apart from $\alpha$. 

It is very surprising that thermodynamics efficiency of such a transport can be very high even 
within anomalous transport regime. This result is not trivial at all. Indeed, the useful work
done by motor in anomalous regime against loading force $f_L$, scales sublinearly in time, 
$W(t)=f_L\langle x(t)\rangle \propto t^{\alpha_{\rm eff} }$ \cite{MMNP13,PRE13,PLoSONE14}. 
However, the free energy 
transformed  into directional motion scales generally as $E_{\rm in}(t)\propto t^\gamma$, where
$0<\gamma \leq 1$.  $\gamma=1$
for rocking, or flashing ratchets driven by either periodic force, or random two state force,
or by random fluctuations of potential characterized by a well-defined mean turnover rate
$\nu=\nu_1\nu_2/(\nu_1+\nu_2)$. Then,  $E_{\rm in}(t)\propto \nu t$. In the energy balance, the
rest, $E_{\rm in}(t)-W(t)$, is dissipated as a net heat $Q(t)$ transferred to the environment.
Thermodynamic efficiency is thus \cite{PhysBio15}
\begin{eqnarray}
R(t)=W(t)/E_{\rm in}(t)\propto 1/t^{\lambda}
\end{eqnarray}
where $\lambda=\gamma-\alpha_{\rm eff}$. Hence, for $R(t)\propto  1/t^{1-\alpha_{\rm eff}}$
for $\gamma=1$. It declines algebraically in time, like also does the mean power 
$P_W(t)=W(t)/t\propto 1/t^{1-\alpha_{\rm eff}}$. However, temporally, for a typical time
required to relocate a cargo within a cell it can be very high, especially when $\alpha_{\rm eff}$ 
is close to one. Even more interesting occurs in the case of bidirectional mechano-chemical coupling,
because the biochemical cycling rates $\nu_{1,2}(x)$ in this case can strongly depend on
 mechanical motion for a sufficiently large $U_0$, when allosteric effects
 start to play a very profound role. Indeed, if available $|\Delta \mu_{\rm ATP}|$ becomes
 smaller than the sum of energies required to enhance the potential energy of motor by two 
 potential
 flashes (see vertical arrow in Fig. \ref{Fig6}) during two halves of one cycle, then the 
 enzyme cycling in its conformational space will not generally stop. It can, however, start to
  occur anomalously slow 
 with a power exponent $\gamma<1$. The average number of forward enzyme turnovers occurring
 with consumption of ATP molecules scales  then as $\langle N(t)\rangle \propto t^\gamma$ in time, and
 $E_{\rm in}(t)=|\Delta \mu_{\rm ATP}|\langle N(t)\rangle \propto t^\gamma$.
 This indeed happens within the model we consider here, see in 
 \cite{PhysBio15} for a particular example with $U_0=30\;k_BT_r\approx 0.75\;eV$, 
 $|\Delta \mu_{\rm ATP}|=20\;k_BT_r\approx 0.5\;eV$, where $\gamma\approx 0.62$
 and $\alpha_{\rm eff}\approx 0.556$ at the optimal load $f_L\approx 8.5$ pN, when the motor pulls
a large cargo at the same time. Thermodynamic efficiency declines in this case very
slowly, with $\lambda\approx 0.067$, so that $R(t)$ is still about 70\% (!) at the
end point of simulations corresponding to physical 3 seconds. Such a high efficiency
is very surprising and should provide one more lesson for our intuition that 
should finally learn and recognize the power of FDT on nanoscale. 
For microscopic and
nanoscopic motion occurring at thermal equilibrium 
the energy lost in frictional processes is regained from thermal random
forces. Therefore, heat losses can, in principle, be small even for an anomalously  strong dissipation.
This is the reason why the attempts to reduce friction on nanoscale are misguided.
They can, pretty counter-intuitively, even hamper efficiency, down to zero as the so-called
dissipationless Brownian (pseudo)-motors reveal.
One should think differently. 

The efficiency at maximum power can also be high in the normal transport operating regime within the
discussed model. Indeed, for $U_0=30\;k_BT_r$ and smaller cargo in \cite{PhysBio15}, the transport
remains almost normal until the maximum of efficiency $R$ is reached at about 80\% for an optimal 
$f_L^{\rm (max)}\sim 9$ pN (see Fig. 8 in \cite{PhysBio15}, where $f_0$ corresponds to $f_L$ here).
 The nearly linear dependence of the efficiency on load  therein
until it reaches about 70\% indicates that the motor steps with almost the same maximal velocity
as at zero loading force $f_L$. 
The following simple heuristic considerations can be used 
to rationalize the numerical results. The motor develops a maximal driving force $F$, which depends
on $U_0$, the motor turnover rate, and temperature (via an entropic contribution), see Fig. 6 in \cite{PCCP14}, where
$F=f_0^{stall}$. It is the stalling 
force. The larger $U_0$, the stronger is the motor and larger $F$. 
 Let us assume that the motor stepping velocity
declines from ${\rm v}_0$ to zero with increasing loading force $f_L$ as 
${\rm v}(f_L)={\rm v}_0[1-(f_L/F)^a]$, where $a\geq 1$ is a power law exponent. Within the 
linear minimalist model of motor considered above, and also 
in an inefficient transport regime within the considered
model, $a=1$, i.e. the motor velocity declines linearly with load. 
However, in a highly efficient nonlinear regime this dependence is strongly nonlinear, $a\gg 1$.
The maximum of 
the motor power $P_W(f_L)=f_L {\rm v}(f_L)$ is reached at $f_L^{\rm (max)}=F/(1+a)^{1/a}$,
with ${\rm v}(f_L^{\rm (max)})=a{\rm v}_0/(1+a)$.
For $a=1$, $f_L^{\rm (max)}=F/2$ and the dependence $P_W(f_L)$ is parabolic. With the
increase of $U_0$, $a$ is also strongly increased and the dependence $P_W(f_L)$ becomes 
strongly skewed, in agreement with numerics. Since the input motor power $P_{\rm in}$
does not depend on load within our model in the energy ratchet regimes, the motor efficiency 
$R=P_W/P_{\rm in}$ just reflects one of $P_W$. Hence, the maximum of $R$ versus $f_L$ does 
correspond  to the maximum efficiency at maximum power and can exceed $1/2$.
The same heuristic considerations can be applied to the results presented 
in Ref. \cite{Parmeggiani} for very efficient normal motors. Of course, these results are
 not necessarily
experimentally relevant for e.g. known kinesin I motors whose maximal efficiency is 
about 50\% \cite{Nelson}. However, 
our theory can be very relevant for devising artificial motors having other tasks because
it provides a biophysically very
 reasonable model where efficiency at maximum power can be larger than the Jacobi bound
of linear stochastic dynamics. It must be stressed, however, that in anomalous transport
regime one cannot define power and one should introduce the notion of sub-power 
\cite{MMNP13,PRE13}.

One should   remark also the following. Even at $f_L=0$, when thermodynamic efficiency is
formally zero, $R(t)=0$, something useful is yet done: the cargo is transferred on a certain distance
by overcoming the dissipative resistance of the environment. 
However, neither potential energy of motor, nor
one of cargo is increased. This is actually normal \textit{modus operandi} of linear molecular
motors like kinesins I or II, very differently from ionic pumps whose the primary goal is to
increase the (electro)chemical potential of transferred ions, i.e. to charge a battery.
Such a $R=0$ regime, should, however, be contrasted with zero efficiency of frictionless
rocking pseudo-ratchets. In our case, the useful work is done against environment. 
Pseudo-ratchets are not capable to do any useful work in principle.

\section{Conclusions}

In this contribution we reviewed some main operating principles of miniscule 
Brownian machines operating on nano- and microscales. Unlike in macroscopic machines, 
thermal fluctuations and noise do play profound and moreover very constructive role
in microworld.
In fact, thermal noise plays a role of stochastic lubricant which supplies energy
to the Brownian machines to compensate for their frictional losses. This is the very
essence of fluctuation-dissipation theorem that the both processes, i.e. frictional
losses and energy gain from the thermal noise are completely compensated on average 
at thermal equilibrium.
Classically, thermal noise vanishes at absolute zero of temperature 
(which physically cannot be achieved anyway, in  accordance with the third law of thermodynamics),
and then only  friction would win (classically). However, quantum noise 
(vacuum zero-point fluctuations) is present 
even at the absolute zero of temperature. Therefore, friction cannot win also at the absolute
zero, and quantum Brownian motion never stops. These fundamental facts allow, in principle,
for a complete transformation of the driver energy into a useful work by isothermal Brownian
engines. Their thermodynamic efficiency approaches unity, when the net heat losses vanish. 
This happens when the motor operates most closely to thermal equilibrium. And this can
be done at any, arbitrarily strong dissipation and  ambient temperatures. No need to
go into a deep quantum cold or to strive for a high quality quantum coherence. A 
most striking example for this is provided by the high transport and thermodynamic
efficiency of molecular motors in subdiffusive transport regime. Operating  anomalously slow, in mathematical
terms, i.e. exhibiting sublinear dependences of both the transport distance and the number of 
motor turnovers on time, such motors can be pretty fast in the absolute terms and work
under a heavy load, see in \cite{PhysBio15}. In this and also other aspects, the intuitive 
understanding  of
subdiffusion and subtransport as extremally slow can be very misleading, 
see in \cite{PRE12b,FNL12,PRL14}. On the other hand, 
the frictionless rocking pseudo-ratchets cannot do any useful work,
as we clarified in this review.

Scientifically sound possibility to approach the theoretical maximum of thermodynamic
 efficiency of isothermal motors
at arbitrarily strong dissipation and ambient temperatures is intrinsically 
related to a possibility of reversible dissipative classical 
computing without heat production.
However, such an adiabatic operation would be infinitesimally slow. Clearly, nobody
needs neither such a motor, nor computer. Moreover, 
adiabatic operation of dissipative pumps involving discrete energy levels is possible
only for a vanishing load. Here, a natural question emerges: What is thermodynamic  efficiency
at maximum power? The linear dynamics result that $R_{\rm max}=1/2$ presents a theoretical 
upper bound is, however,
generally completely wrong within nonlinear stochastic dynamics, as we showed
in this review with three (!) examples. This opens a door for
design of highly efficient Brownian and molecular motors.
Moreover, the recent model results in \cite{PhysBio15} for  a normal transport of sufficiently
small subdiffusing (when free) cargos by a kinesin motor with a very high thermodynamical 
efficiency at optimal external load do imply that thermodynamical efficiency at maximum power within
that model can also be well above 50\%. The earlier results for a normal 
diffusion molecular motors within a model of a very similar kind \cite{Julicher,AstumianBier} 
obtained in \cite{Parmeggiani} also corroborate such conclusions. Such models are able
to mimic allosteric interactions within minimalist model setups. Chemical allosteric
interactions, which are intrinsically highly nonlinear, can optimize performance of various 
molecular motors. This line of reasoning is especially important for design of 
artificial molecular motors \cite{Cheng}. It should be pursued further.
 Quantum effects can also be important
to arrive at this goal, even when quantum coherence does not play any role, i.e.
on the level of rate dynamics with quantum rates (like in Pauli master equation).
In particular, it has been shown in this review within a simplest toy model possible
that quantum effects (related to inverted regime of quantum particle transfer) 
can lead to thermodynamic efficiencies at maximal power larger than one half for the machine
operating both in direct and reverse directions.
Quantum coherence can also play here  a role, which should be clarified in further research.
Undoubtedly, quantum coherence is central for quantum computing, which is obviously
reversible \cite{Feynman}. However, this is a different story.

I hope that the readers of this review will find it most useful in, especially, liberating
themselves and possibly others from
some common fallacies, both spoken and unspoken 
which unfortunately pervaded literature and hinder the further progress. A valid coherent
picture emerges.



\begin{acknowledgements}
Support of this research by the Deutsche Forschungsgemeinschaft 
(German Research Foundation), Grant GO 2052/1-2 is gratefully acknowledged.
\end{acknowledgements}





\bibliography{review}

\end{document}